\documentclass[twocolumn,showpacs,preprintnumbers,amsmath,amssymb,floatfix,nofootinbib,linenumbers]{revtex4}
\usepackage{color}
\usepackage{graphicx}
\usepackage{dcolumn}
\usepackage{bm}
\usepackage{slashed}
\usepackage{float}
\usepackage{changepage}
\usepackage{ulem} 
\usepackage{graphicx}
\usepackage[utf8]{inputenc}
\usepackage{natbib}
\usepackage{subcaption}
\usepackage{caption}

\begin{document}
\def\bb#1{\hbox{\boldmath${#1}$}}

\title{ Mini-jet clustering algorithm using transverse-momentum seeds in high-energy nuclear collisions}

\author{Hanpu Jiang$^{1,2}$, Nanxi Yao$^{1,3}$, Cheuk-Yin Wong$^4$, Gang Wang$^1$, Huan Zhong Huang$^{1,5}$}

\affiliation{$^1$Department of Physics and Astronomy, University of California, Los Angeles, California 90095, USA}

\affiliation{$^2$Department of Physics, Columbia University in the City of New York, New York, New York 10027, USA}

\affiliation{$^3$Department of Physics, University of Illinois at Urbana-Champaign, Champaign, Illinois 61801, USA}

\affiliation{$^4$Physics Division, Oak Ridge National Laboratory\footnote{
    This manuscript has been authored in part by UT-Battelle, LLC, under contract DE-AC05-00OR22725 with the US Department of Energy (DOE). The US government retains the publisher, by accepting the article for publication acknowledges that the US government retains a nonexclusive, paid-up, irrevocable, worldwide license to publish or reproduce the published form of this manuscript, or allow others to do so, for US government purposes. DOE will provide public access to these results of federally sponsored research in accordance with the DOE Public Access Plan (http://energy.gov/downloads/doe-public-access-plan), Oak Ridge, Tennessee 37831, USA }\!, Oak Ridge, Tennessee 37831, USA}

\affiliation{$^5$Key Laboratory of Nuclear Physics and Ion-beam Application (MOE) and Institute of Modern Physics,\\Fudan University, Shanghai 200433, China\\}

\date{\today}

\begin{abstract} 
{We propose an algorithm to detect mini-jet clusters in high-energy nuclear collisions, by selecting a high-transverse-momentum ($p_T$) particle as a seed and assigning a clustering radius ($R$) in the pseudorapidity and azimuthal-angle space. Our PYTHIA simulations for $p$+$p$ collisions show that a scheme with a seeding $p_T$ of around 0.5 GeV/$c$ and $R$ of approximately 0.6 satisfactorily identifies mini-jet clusters. The correlation between clusters obtained in PYTHIA calculations using the algorithm exhibits the proper behavior of hard-scattering-like processes, suggesting its usefulness in isolating mini-jet-like clusters from non-hard-scattering soft processes when applied to actual nuclear-collision data, thereby allowing a closer examination of both the mini-jet and the soft mechanisms.}
\end{abstract}


\maketitle

\section{Introduction}

The hard-scattering process was originally proposed as the primary mechanism for producing high-transverse-momentum ($p_T$) jet clusters, typically in the range of tens of GeV/$c$ \cite{Bla74, Ang78, Fey78, Owe78, Rak13, Sjo86, Sjo87}. However, the UA1 Collaboration found that the production of particle clusters with a total $p_T$ of a few GeV/$c$ is an important production process in hadron-hadron collisions, representing a sizeable fraction of the particle production cross-sections \cite{UA1}. Jet clusters in such a $p_T$ range were coined ``mini-jet"~\cite{Esk89}. The dominance of jet production extends to even lower-$p_T$ domains at higher collision energies ($\sqrt{s}$) because (i) the fraction of particles produced by such a process increases rapidly with $\sqrt{s}$, and (ii) the jet-production invariant cross-section at midrapidities varies inversely with $p_T$ \cite{Wan91,Ada06,Esk89,Sjo06}. In fact, the mini-jet dominance has been found to extend to the $p_T$ region of a few tenths of a GeV/$c$ in high-energy $p$+$p$ and $p$+$\bar{p}$ collisions at $\sqrt{s}$ = 0.9 to 7 TeV \cite{Won13, Won15}.
 
While the pQCD-type hard-scattering process remains a key source of particle production, soft processes contribute to the majority of low-$p_T$ particles with distinct outcomes such as rings, pancakes, granular droplets, hydrodynamical collective motions, quark-gluon plasma, and color-glass condensate \cite{Bjo83, Ste07, Hwa74, Lan53, Bel56, Ama57, Car73, Coo74, Cha74, Cse87, Sri92, Ham01, Mcl94, Mcl94a, Ian02, Ian03, Gel10, Moh03, Pra07, Bia07, Osa09, Sar06, Won74, Bas81, Cse92, Mag01, Won04, Zha05, Cso08, Beu08, Tra14}. To better understand the underlying reaction processes for such production mechanisms, it is necessary to devise a method to distinguish between the part produced by mini-jet clusters and the part produced by soft processes. Isolating the mini-jet clusters by such a method will not only benefit the study of the mini-jet dynamics but also help to identify the soft part for further examination.

Previously, as a first step to identify mini-jets, we studied the clustering properties of produced particles in high-energy $p$+$p$ collisions in the space of pseudorapidity ($\eta$) and azimuthal angle ($\phi$). We devised an algorithm to find mini-jet-like clusters by using the $k$-means clustering method in conjunction with a $k$-number (cluster number) selection principle \cite{Won20}. To study the clustering algorithm, we examined minimum-bias events of $p$+$p$ collisions at $\sqrt{s}$ = 200 GeV generated by PYTHIA8.1~\cite{Sjo07}. Our findings indicate that multiple mini-jet-like and mini-dijet-like clusters of low-$p_T$ hadrons can arise in high-multiplicity events. However, comparable clustering behavior is also evident for randomly produced particles confined to a finite $\eta$ and $\phi$ space. Therefore, the ability to discern azimuthally back-to-back correlated mini-jet-like clusters as bona fide mini-jets and mini-dijets will rely on further correlation requirements.

A mini-jet, as a cluster of particles originating from a high-$p_T$ jet cascade, is likely to contain some high-$p_T$ particles. Thus, it is reasonable to use a high-$p_T$ particle as a seed for identifying mini-jets. To fix the mini-jet cluster number $K$ and also to eliminate the noise that may contaminate the mini-jet clusters, we add a supplementary condition that requires each mini-jet to contain at least one seed particle with $p_T$ greater than a certain threshold value, $p_{T0}$. The fixed cluster number and the knowledge of the mini-jet size in $\eta$ and $\phi$ bring us closer to understanding the mini-jet and their correlations. The remaining particles can be attributed to non-jet and non-mini-jet processes, which form the bulk part of the dynamics. Clearly, the higher  $p_{T0}$, the more the mini-jet cluster will resemble the conventional high-$p_T$ jet, and the higher the chance of detecting its dijet partner. However, we focus on mini-jets in the very low-$p_T$ domain because they are more abundant in the underlying events. We analyze the clustering properties of mini-jets as a function of the seed $p_{T0}$ value.

To ensure that the reconstructed clusters are mini-jet clusters, we rely on the inter-cluster correlations.
Analytical expressions for the hard scattering process summarize its essential features and dependencies, making it easier to uncover dynamical effects wherever they may occur. These analyses have revealed insights into the dominance of the hard-scattering process in the high-$p_T$ domain and have helped to locate the boundary between the hard-scattering and flux-tube fragmentation processes in $p$+$p$ collisions at high energies \cite{ Won15}.

In Section~\ref{Sec:II}, we first derive the differential cross-section $E_c E_\kappa d\sigma (AB \to c\kappa X) /d\bb c \,d \bb \kappa$ for the production of two massive partons $c$ and $\kappa$. This allows us to integrate out $p_T$ to acquire the two-particle correlation function $d\sigma/d\Delta \phi \,d\Delta y$ in Section~\ref{Sec:III}, where $\Delta \phi =\phi_{\kappa}-\phi_c$ and $\Delta y=y_{\kappa}- y_c$. This correlation function exhibits the "ridge" structure on the away side at $\Delta \phi\approx \pm \pi$ in the hard-scattering process.
Next, in Section~\ref{Sec:IV}, we present the new clustering algorithm and demonstrate its application to PYTHIA data of $p$+$p$ collisions.
In Section~\ref{Sec:V}, we compare the two-cluster correlations obtained from the algorithm with the two-particle correlations. Additionally, in Section~\ref{Sec:VI}, we discuss how the anti-$k_T$ algorithm complements our new clustering algorithm. Finally, we draw brief conclusions in Section~\ref{Sec:VII}.

\section{Hard Scattering Integral for $E_c E_\kappa d\sigma(AB\to c\kappa X) /d\bb c \,d\bb \kappa$ }
\label{Sec:II}

In the parton model, the hard-scattering cross-section for $AB\to c \kappa X$, the production of partons $c$ and $\kappa$ with momenta $c$ and $\kappa$,
is given by \cite{Owe87,Won16}
\begin{eqnarray}
d\sigma ( AB &\to& c \kappa X) =\sum_{ab} \int K_{ab} dx_a d{\bb a}_T dx_b d{\bb b}_T 
\nonumber\\	
&\times& G_{a/A}(x_a,{\bb a}_T) G_{b/B} (x_b,{\bb b}_T) d\sigma( ab \to c \kappa) ,~~
\end{eqnarray}
where $(x_a, \bb a_T) $ and $(x_b, \bb b_T) $ represent the momenta, and $G_{a/A}$ and $G_{b/B}$, the structure functions of the incident partons $a$ and $b$, respectively. $K_{ab}$ is a correction factor that can be obtained perturbatively~\cite{Sjo15}
or approximated nonperturbatively~\cite{Cha95}. The quantity $d\sigma( ab \to c \kappa)$ is the cross-section element for the process $ab \to c\kappa$,
\begin{eqnarray}
d\sigma (ab&\to& c\kappa)= \frac{ 1 }{4[(a \cdot b)^2 - m_a^2 m_b^2 ] ^{1/2}} |{ T}_{fi}|^2
\nonumber\\
&\times& \frac{d^3c}{(2\pi)^3 2 E_ c} \frac{ d^3\bb \kappa}{(2\pi)^3 2 E_ {\kappa}} (2 \pi)^4 \delta^4 ( a+b- c-\kappa).~~~~
\end{eqnarray}
Here, we normalize the Dirac fields by $\bar u u = 2m$. The quantity $|T_{fi}|^2$ is related to $d\sigma/dt$ by
\begin{eqnarray}
|T_{fi}|^2 = 16 \pi [\hat s-(m_a+m_b)^2][\hat s-(m_a-m_b)^2] \frac{d\sigma(ab\to c\kappa) }{dt}.
\label{8}
\nonumber
\end{eqnarray}
We consider the simplified case with $m_a=m_b=0$ and treat $a_T$ and $b_T$ as small perturbations. The cross-section element is then
\begin{eqnarray}
d\sigma(ab\to c\kappa) = \frac{ s_{ab}}{2\pi}\! \frac{d\sigma(ab\to c\kappa) }{dt} \frac{d^3c}{ E_ c} \frac{d^3\kappa}{ E_ \kappa} \delta^4 ( a+b - c - \kappa),
\end{eqnarray}
where $\hat s=s_{ab}=(a+b)^2$, which is different from $s=s_{AB}=(A+B)^2$. We obtain
\begin{eqnarray}
&&\hspace*{-0.6cm}
\frac{E_c E_\kappa d\sigma (\! AB\!\! \to\!\!c \kappa X\!)}{d^3 c ~ d^3 \kappa} \!=\!\sum_{ab}\!\!\! \int \!\!K_{ab} dx_a d{\bb a}_T dx_b d{\bb b}_T 
\nonumber\\
&&\hspace*{-0.6cm}\times G_{a/A}\!(x_a,\!{\bb a}_T\!) G_{b/B}\! (x_b,\!{\bb b}_T\!) \!\frac{\hat sd\sigma \!(ab\!\to\! c\kappa) }{2\pi dt} \delta^4\! ( a\!+\!b\! -\!c \!-\! \kappa).~
\label{eq5}
\end{eqnarray}
We consider a factorizable structure function with a Gaussian intrinsic transverse momentum distribution,
\begin{eqnarray}
G_{a/A}(x_a,{\bb a}_T)=G_{a/A}(x_a) \frac{1}{2\pi \sigma^2} e^{-{\bb a}_T^2/2\sigma^2},
\end{eqnarray}
where $2\sigma^2 \sim 0.9 $ (GeV/$c$)$^2$ \cite{Won98}.
Upon integrating over the transverse momenta $\bb a_T$ and $\bb b_T$, we reach
\begin{eqnarray}
&&\hspace*{-0.5cm}\frac{d\sigma ( AB \to c \kappa X)}{dy_c c_T dc_T d \phi_c\, dy_\kappa \kappa_T d\kappa_T d \phi_\kappa } =\sum_{ab} \int K_{ab} dx_a dx_b
\nonumber\\
&&\times G_{a/A}(x_a)G_{b/B} (x_b) \frac{ e^{-\frac{(\bb c_T + \bb \kappa_T)^2}{4\sigma^2}}} {2(2\pi \sigma^2)} \frac{ \hat s}{2\pi}\frac{d\sigma(ab\to c\kappa) }{dt} \nonumber\\
& &\times \delta ( a_0+b_0 - (c_0+\kappa_0))\delta ( a_z+b_z - (c_z+\kappa_z)).
\label{eq8} 
\end{eqnarray}
To perform the integration over $x_a$ and $x_b$, we spell out the momenta in the infinite momentum frame,
\begin{eqnarray}
a&=&(x_a \frac{\sqrt{s}}{2} + \frac{a^2+a_T^2}{2x_a \sqrt{s}}, ~{\bb a}_T, ~~x_a \frac{\sqrt{s}}{2} - \frac{a^2+a_T^2}{2x_a \sqrt{s}}),\\
b&=&(x_b \frac{\sqrt{s}}{2} + \frac{b^2+b_T^2}{2x_b \sqrt{s}}, ~\, {\bb b}_T,-x_b \frac{\sqrt{s}}{2} + \frac{b^2+ b_T^2}{2x_b \sqrt{s}}),\\
c&=&(x_c \frac{\sqrt{s}}{2} + \frac{c^2+c_{T}^2}{2x_c \sqrt{s}}, ~{\bb c}_T, ~~x_c \frac{\sqrt{s}}{2} - \frac{c^2+c_T^2}{2x_c \sqrt{s}}) ,\\
\kappa&=&(x_\kappa \frac{\sqrt{s}}{2} + \frac{\kappa^2+\kappa_T^2}{2x_\kappa \sqrt{s}},\,{\bb \kappa}_T, -x_\kappa \frac{\sqrt{s}}{2} + \frac{\kappa^2+\kappa_T^2}{2x_\kappa \sqrt{s}}),
\end{eqnarray}
where $x_c$ and $x_\kappa$ can be represented by $y_c$ and $ y_\kappa$, respectively 
\begin{eqnarray}
x_c =\frac{m_{cT} e^{y_c} }{\sqrt{s}}, ~~~ x_\kappa = \frac{m_{\kappa T} e^{y_\kappa} }{\sqrt{s}}.
\end{eqnarray}
The two $\delta$ functions in Eq.\ (\ref{eq8}) can be integrated to yield
\begin{eqnarray}
& &
\frac{d\sigma ( AB \to c d X)}{dy_c c_T dc_T d \phi_c\, dy_\kappa \kappa_T d\kappa_T d \phi_\kappa }
\nonumber \\ &=&
\sum_{ab} K_{ab} x_{a}G_{a/A}(x_{a}) x_{b}G_{b/B} (x_{b})
 \frac{ e^{-\frac{(\bb c_T + \bb \kappa_T)^2}{4\sigma^2}}} {2\pi (4\pi \sigma^2)}
\nonumber\\
& &\times  \frac{d\sigma(ab\to c\kappa) }{dt},
\label{eq12} 
\end{eqnarray}
where
\begin{eqnarray}
x_a&&= x_c + \frac{\kappa^2+\kappa_T^2}{x_\kappa s } - \frac{b^2+ b_T^2}{x_b s} 
\nonumber\\
&&= \frac{m_{cT} e^{y_c} }{\sqrt{s}} + \frac{m_{\kappa T} e^{-y_\kappa } }{\sqrt{s}} - \frac{b^2+b_T^2}{x_b s}, \nonumber\\
x_b&& =x_\kappa +\frac{c^2+c_{T}^2}{x_c {s}}-\frac{a^2+a_T^2}{x_a s}
\nonumber\\
&& =\frac{m_{\kappa T}e^{y_\kappa}}{\sqrt{s}} +\frac{m_{cT}e^{ -y_c} }{\sqrt{s}} -\frac{a^2+a_T^2}{x_a s}.
\end{eqnarray}
The above formula renders the cross-section for the production of $c$ and $\kappa$, when the elementary cross-section ${d\sigma(ab\to c\kappa) }/{dt}$ is given explicitly in terms of its dependent variables.

\section{ The angular correlation $d\sigma(AB \to c\kappa X)/d\Delta \phi \,d\Delta y$}
\label{Sec:III}

To obtain the angular correlations between particles $c$ and $\kappa$, we transform $(y_c, y_\kappa, \phi_c, \phi_\kappa )$ 
  to $(Y,\Delta y, \Phi, \Delta \phi)$,
\begin{eqnarray}
y_c = Y + \Delta y/2, && y_\kappa = Y-\Delta y/2,
\nonumber \\	
\phi_c = \Phi + \Delta \phi/2, && \Phi_\kappa = \Phi-\Delta \phi /2.
\end{eqnarray}
We then have 
\begin{eqnarray}
dy_c\, dy_\kappa \, d \phi_c\, d\phi_\kappa = dY\, d\Delta y \, d\Phi \, \Delta \phi.
\end{eqnarray}
Equation (\ref{eq12}) becomes
\begin{eqnarray}
&&\hspace*{-0.3cm}\frac{d\sigma ( AB \to c \kappa X)}{ d\Delta \phi d\Delta y d Y d\Phi }= \sum_{ab} K_{ab} \int x_{a}G_{a/A}(x_{a}) x_{b}G_{b/B} (x_{b})~~~~
\nonumber\\
 & &\hspace*{0.3cm}\times c_T dc_T \kappa_T d\kappa_T 
 \frac{ e^{-\frac{(\bb c_T + \bb \kappa_T)^2}{4\sigma^2}}} {2\pi (4\pi \sigma^2)} \frac{d\sigma(ab\to c\kappa) }{dt}.
\label{eq16} 
\end{eqnarray}
The cross-section should be independent of the average value of $\phi_c$ and $\phi_\kappa$. The integration over $\Phi$ gives 
\begin{eqnarray}
&&\hspace*{-0.3cm}\frac{d\sigma ( AB \to c \kappa X)}{ d\Delta \phi d\Delta y d Y }= \sum_{ab} K_{ab} \int c_T dc_T \kappa_T d\kappa_T ~~~~
\nonumber\\
 & &\hspace*{-0.3cm}\times 
x_{a}G_{a/A}(x_{a}) x_{b}G_{b/B} (x_{b})
 \frac{ e^{-\frac{(\bb c_T + \bb \kappa_T)^2}{4\sigma^2}}} {(4\pi \sigma^2)} \frac{d\sigma(ab\to c\kappa) }{dt}.
\label{eq17} 
\end{eqnarray}

We  focus on $Y=0$ and 
consider approximate boost invariance at midrapidities. The correlation function Eq.~(\ref{eq12}) from the process $ab$$\to$$c\kappa$
 at midrapidities becomes
\begin{eqnarray}
\!\frac{d\sigma ( AB \to c \kappa X)}{ d\Delta \phi \, d\Delta y }\bigg |_{Y=0} \!= \!K_{ab} C( \Delta \phi, \Delta y),~~
\label{eq14}
\end{eqnarray}
where
\begin{eqnarray}
&&
\hspace*{-0.5cm}C( \Delta \phi,\Delta y ) = \int_0^\infty \!\!c_T dc_T \, \int _0^\infty\!\! \kappa_T d\kappa_T x_a G_{a/A}(x_{a}) x_b G_{b/B} (x_{b})
 \nonumber\\
&&\hspace*{-0.8cm}\times \frac{1} { (4\pi \sigma^2)} \exp\{-\frac{c_T^2 + 2 c_T \kappa_T \cos \Delta \phi + \kappa_T^2}{4\sigma^2}\} \frac{d\sigma(ab \to c{\kappa}) }{dt}.
\label{eq19}
\end{eqnarray}

The basic cross-section $d\sigma /dt$ can be written in terms of relativistic invariant quantities ($s', t', u'$) measured in the intrinsic system of $a + b \to c +\kappa$. We have Eq.~(10.169) in Ref.~\cite{Gas88} for $gg \to gg$, 
\begin{eqnarray}
\frac{d\sigma(g g \to gg ) }{dt}= \frac{9 \pi\alpha_s^2}{8} \frac{(s'^4+t'^4+u'^4)(s'^2+t'^2+u'^2)}{s'^4t'^2u'^2}.
\label{eq20}
\end{eqnarray}
From Eq.~(2.12) of Ref.~\cite{Com79} for $gg \to q \bar q$, and Eq.~(A1) of
Ref.~\cite{Glu78}, 
 with quark mass $m_c=m_\kappa=m$, we have for $gg \to q \bar q$,
\begin{eqnarray}
\frac{d\sigma}{dt}(gg\to q \bar q) =&&\frac{ \pi \alpha^2}{16 s'^2} \biggr [ \frac{12}{s^2}(m^2-t')(m^2-u') 
\nonumber\\
&&+ \frac{8}{3} \frac{(m^2-t')(m^2-u')-2m^2(m^2+t')}{(m^2-t')^2}
\nonumber\\
&&
+ \frac{8}{3} \frac{ (m^2-t')(m^2-u')-2m^2(m^2+u')}{(m^2-u')^2}
\nonumber\\
&&-\frac{2m^2(s'-4m^2)}{3(m^2-t')(m^2-u')}
\nonumber\\
&&
-6\frac{(m^2-t')(m^2-u')+m^2(u'-t')}{s'(m^2-t')}
\nonumber\\
&&-6\frac{(m^2-t')(m^2-u')+m^2(t'-u')}{s'(m^2-u')}\biggr ].~~~
\label{86}
\end{eqnarray}
From Eq.~(2.7) of Ref.~\cite{Com79}, we have for $q\bar q \to c \bar c$
\begin{eqnarray}
\frac{d\sigma (q\bar q \to c \bar c)}{dt} = \frac{4}{9}\frac{ \pi \alpha^2}{s'^2} \frac{(m^2 - t')^2 +(m^2-u')^2 + 2m^2 s'} {s'^2}.~~
\end{eqnarray}

The structure function can be expressed in the form $ x_a G_{a/A}(x_{a}) $$\propto $$(1-x_a)^{g_a} $,
for which the two-particle angular correlation function becomes approximately 
\begin{eqnarray}
\frac{d\sigma ( AB \!\to \!c \kappa X)}{ d\Delta \phi \, d\Delta y} \bigg | _{Y\sim 0} \! \!\!&&\hspace*{-0.4cm} \sim \! A\! \biggl [ \!1 \!- \!\frac{4m_{cT}}{\sqrt{s}} \cosh \Delta y + O((\frac{m_{cT}}{\sqrt{s}})^2)\biggr]^{g_a}
\nonumber\\
&&\times C( \Delta \phi).~~
\end{eqnarray}
This indicates that as $m_{cT}/\sqrt{s'} $ is much less than unity, the correlation function is a weak function of $\Delta y$, and 
\begin{eqnarray}
&&
\hspace*{-0.5cm}C( \Delta \phi ) = \int_0^\infty \!\!c_T dc_T \, \int _0^\infty\!\! \kappa_T d\kappa_T 
\nonumber\\
&&\hspace*{-0.8cm}\times \frac{1} { (4\pi \sigma^2)} \exp\{-\frac{c_T^2 + 2 c_T \kappa_T \cos \Delta \phi + \kappa_T^2}{4\sigma^2}\} \frac{d\sigma(ab \to c{\kappa}) }{dt}.
\label{eq24}
\end{eqnarray}

\begin{figure}[H]
\centering 
\includegraphics[width = .99\linewidth]{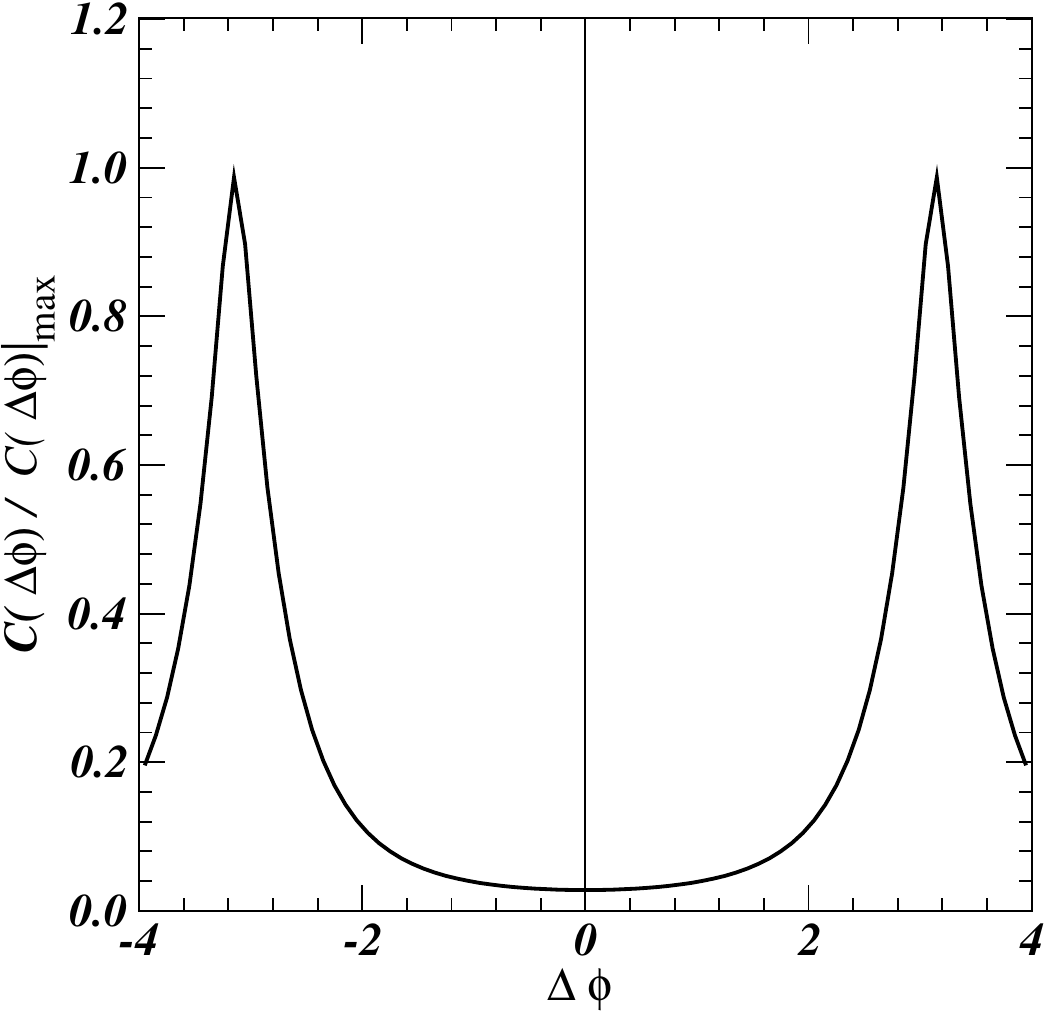}
\caption{The correlation function $C(\Delta \phi)$.}
\end{figure}

From Eqs.~(\ref{eq19}) and (\ref{eq24}), we expect that the contribution to the correlated production of particles $c$ and $\kappa$ is the greatest when \textcolor{black}{$\Delta \phi=\pm \pi$}, for which particles $c$ and $\kappa$  are produced positioned back-to-back. 
There is no completely rigorous model to go from pQCD for produced particles in the low-$p_T$ region because the mechanism of the non-perturbative hadronization is not fully known. One relies on semi-empirical fragmentation functions to describe the fragmentation of the partons to hadrons $c$ and $\kappa$ in the hard-scattering model.

To grasp the gross feature of the correlation function, we use the simple parton-hadron duality and approximate $d\sigma(ab$$ \to$$ c\kappa)/dt$ to be of the form as in $gg \to gg$ collisions in Eq.~(\ref{eq20}). We have
\begin{eqnarray}
&&s' \sim 4m_{cT}^2\cosh^2 Y,
\nonumber\\
&&t'-m_c^2
=- 2\cosh Y e^{-Y} m_{cT}^2 =(1+ e^{-2Y}) m_{cT}^2,
\nonumber\\
&&
 u'-m_c^2= 
 - 2\cosh Y e^{Y}~ m_{cT}^2 = - (1+ e^{2Y})m_{cT}^2. ~~~~
\end{eqnarray}
For $gg \to gg$ with small $m_{cT}$, we have $d\sigma(gg \to gg)/{dt}\sim 1/m_{cT}^4$, and 
\begin{eqnarray}
\frac{d\sigma(ab \to c\kappa)}{dt} \sim \frac{1}{t'^2} \sim \frac{A'}{ [1+ (m_c^2+c_T^2)/M_{T0}^2 ]^{n/2}},  
\end{eqnarray}
where $n \sim 4$ in pQCD and $M_{{}{T0}}$ is the regularizing mass 
on the order of 1 GeV/$c^2$ empirically in the hadron-parton duality approximation~\cite{Won15}.
The numerical integration over $c_T$ and $\kappa_T$ in Eq.\ (\ref{eq24}) gives the correlation function $C(\Delta \phi)$ as shown in Fig.~1. 
The correlation function for $m_{\rm parton}=0.14$ GeV/$c^2$ has maxima at $\Delta\phi \approx\pm \pi$ and a minimum at $\Delta \phi=0$ because of the elementary back-to-back production process of $a+b \to c +\kappa$. It is relatively flat in $\Delta y$ because $m_{cT}/\sqrt{s} \ll 1$ in Eq.\ (\ref{eq16}) for high-energy collisions. This gives the feature of the correlation function in the form of a ridge structure on the away side at $\Delta\phi\approx \pm \pi$, with a back-to-back azimuthal correlation in $\phi$, 
as expected from momentum conservation.

\section{$p_T$-seeded Clustering Algorithm}
\label{Sec:IV}

Because the mini-jet manifests as a concentrated release of energy with a high-$p_T$ value during its parton-cascading evolution, some products of its on-shell hadronized particles are likely to retain high $p_T$ at the endpoints of the evolution. For this reason, we assume that every mini-jet has a high-$p_T$ remnant. The magnitude of the $p_T$ value, which allows such a characterization, will need to be determined semi-empirically. The mini-jet is also expected to evolve in a certain direction, with the cluster of its evolution products likely within a bundle in the pseudorapidity angles and azimuthal angles. Therefore, we develop the $p_T$-seeded clustering algorithm to idealize mini-jet as a cluster of hadrons in the $(\eta,\phi)$ space with a radius previously estimated to be on the order of $R=0.5-0.6$ \cite{Won09}. For a given set of $M$ particles in an event with specified positions $x_i=(\eta_i,\phi_i)$ and transverse momenta $p_{Ti}$, where $i=1,2,...M$ is the particle index, particles above a threshold $p_{T0}$ are selected as the seed particles. These seed particles serve as indicators of mini-jets and determine the initial number and locations of the clusters, while the low-$p_T$ particles are used for the cluster location refinement in later steps. Particles outside the mini-jet cones are regarded as products of non-jet processes.

Each seed particle is initially treated as the center of a seeded cluster. To avoid overlapping, we merge clusters that are sufficiently close to each other to form a single cluster. \textcolor{black}{The merging process calculates the distance between each pair of centers, $\ell=[(\Delta \eta)^2 + (\Delta \phi)^2]^{1/2}$, and finds the two closest centers. If this shortest distance is less than a certain threshold, 0.816 in this algorithm, we merge the two clusters. With a minijet modeled as a cone with a given radius $R= 0.6$, 
we envisage that two seed candidates 
will have an 83.4\% likelihood of belonging to the same minijet if their $\ell$ is less than 0.816. The assigned value of merging threshold 0.816 is derived from the hypothetical scenario wherein two points are randomly sampled from inside a circle with a radius of $R=0.6$, and the distance between them adheres to $\rho(\ell)$, the distance distribution in a uniform disk~\cite{Pel05}:
\begin{equation}
\rho(\ell) = \frac{4\ell}{\pi R^2} \arccos\left(\frac{\ell}{2R}\right) - \frac{2\ell^2}{\pi R^4} \sqrt{R^2 - \frac{\ell^2}{4}}.
\end{equation}
}After calculating the mean $\mu$ and the width $\sigma$ of the distance distribution, we obtain $\mu + \sigma \approx 0.816$. The range [0, 0.816] corresponds to approximately 83.4\% probability of the distance distribution. The cumulant probability can be increased to 90\%(95\%) with the upper bound of 0.894(0.974). 
\textcolor{black}{Various merging thresholds can be explored by examining the physics of the jet cascade, for example, in PYTHIA.}
The new center of the merged cluster is calculated as the weighted midpoint of the previous two seed particles, with the weight assigned based on their $p_T$. This merging process continues until the distance between any two clusters is greater than 0.816. At this point, the cluster positions are roughly determined, based on the seed particles.

\begin{figure}[]\centering
\includegraphics[width = .99\linewidth]{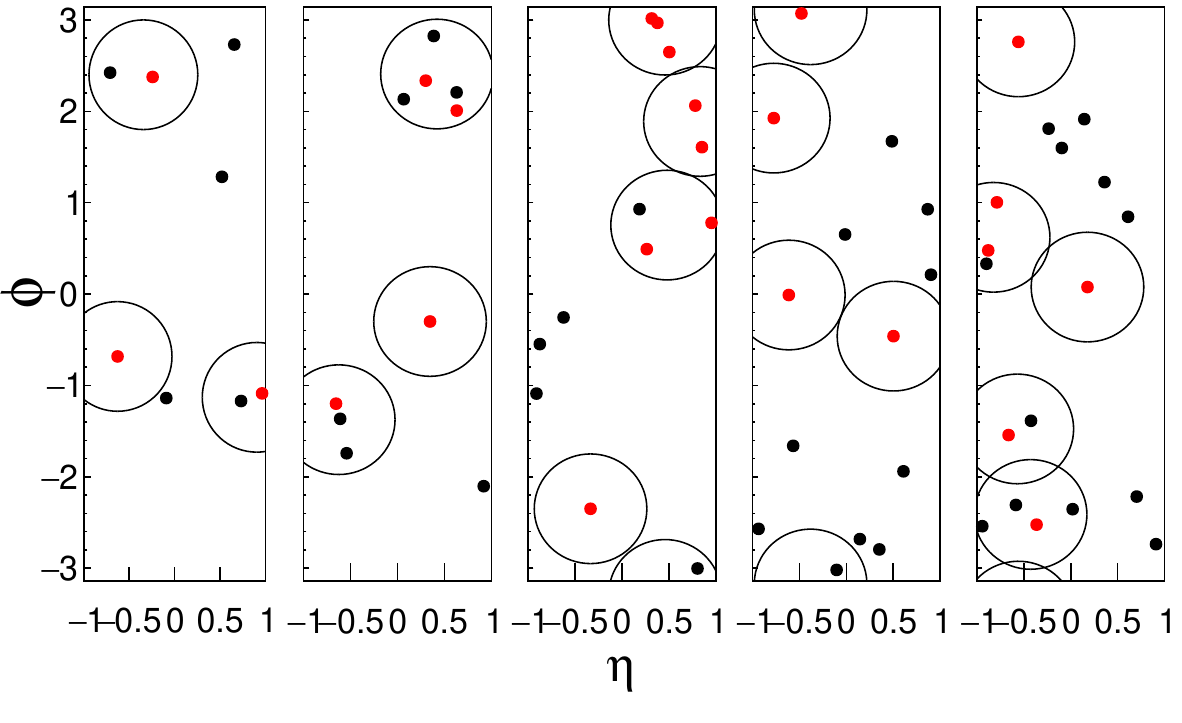}
\caption{Examples with $p_{T0} = 0.5$ GeV/$c$. The seed particles are shown as red points.}
\label{fig0-0}\end{figure}

To enhance the precision and reliability of the cluster positions, all particles are used to recalculate the cluster centers as a refinement. The final cluster centers are calculated using a formula:
\begin{eqnarray}
 {\bb C}_k=\frac { \sum_{{\bb x}_i^k \in S_k} \left(\frac{p_{Ti}}{\langle p_T\rangle}\right)^\gamma {\bb x}_i^{k} } 
 {\sum_{{\bb x}_i^k \in S_k} \left(\frac{p_{Ti}}{\langle p_T\rangle}\right)^\gamma } ,
\end{eqnarray}
\label{eqdisk}
where the subscript $k$ represents the cluster index, and $S_k$ denotes the subset of particles located within a radius of 0.6 around the center of the corresponding seeded cluster. A power index of $\gamma=2$ is used in the algorithm. The resultant clusters are finally settled after the refinement of cluster positions.

Figure~\ref{fig0-0} illustrates examples of the $p_T$-seeded clustering algorithm using a threshold of $p_{T0} =0.5$ GeV/$c$. The high-$p_T$ particles, denoted by red points, are initially assigned as independent cluster centers. Seeds that are close to each other are merged, as shown in the upper right corner of the third panel. The center positions of clusters are then refined by the surrounding low-$p_T$ particles. In cases where no other particles are present around a seed particle, a cluster is still formed with the seed particle as the center, as demonstrated in the second and fourth panels.

\section{Correlations Between Particles and Between Clusters}
\label{Sec:V}

\subsection{Two-Particles Correlations}

\begin{figure}[tbhp]\centering
\includegraphics[width = .99\linewidth]{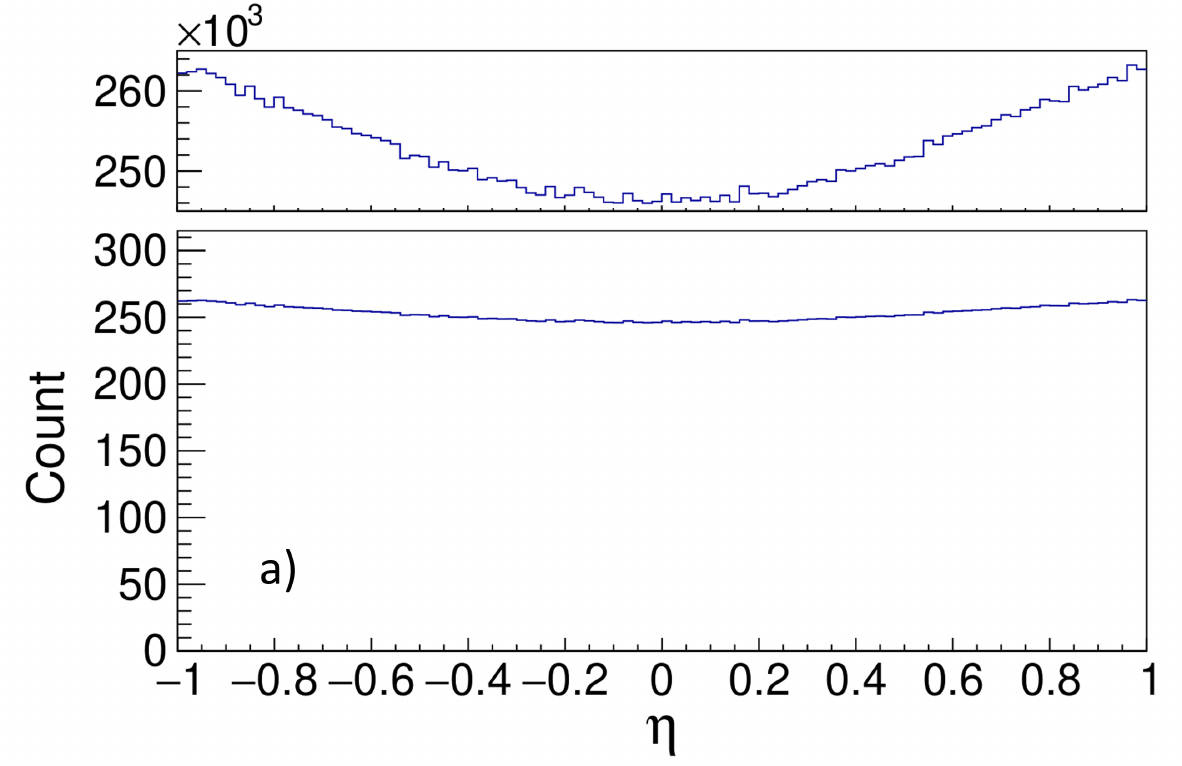}
\includegraphics[width = .99\linewidth]{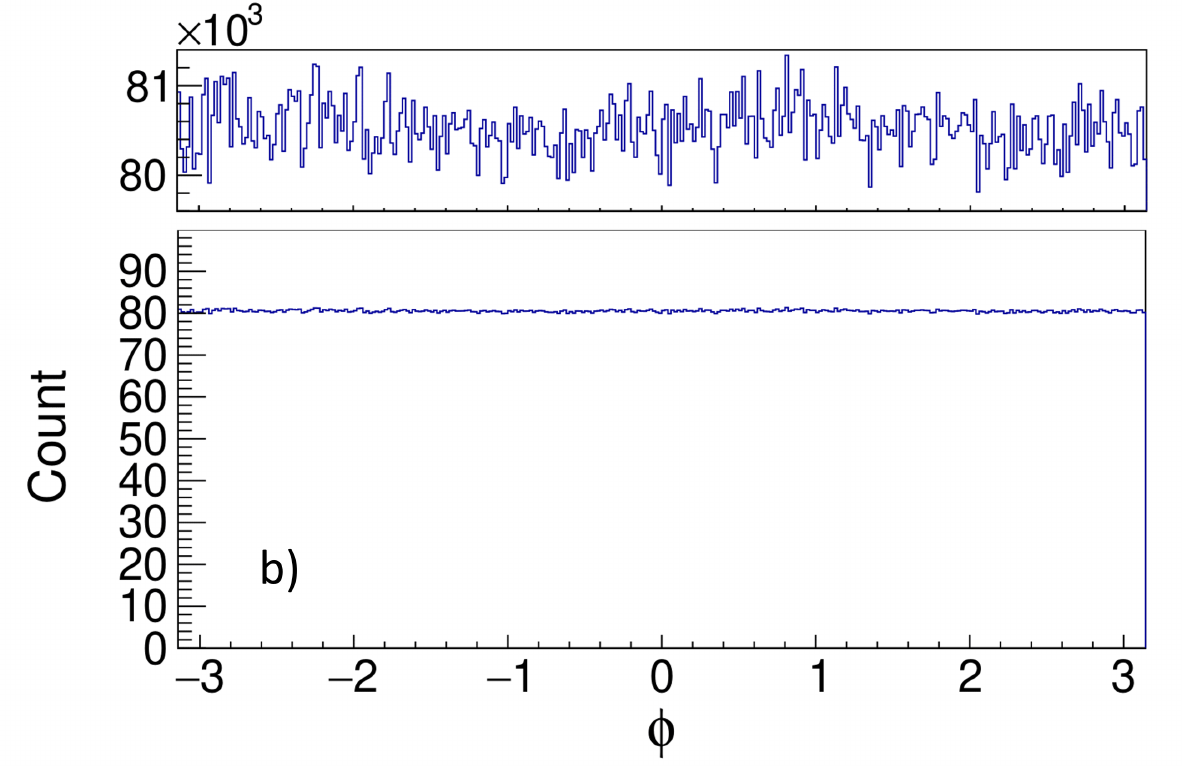}
\includegraphics[width = .99\linewidth]{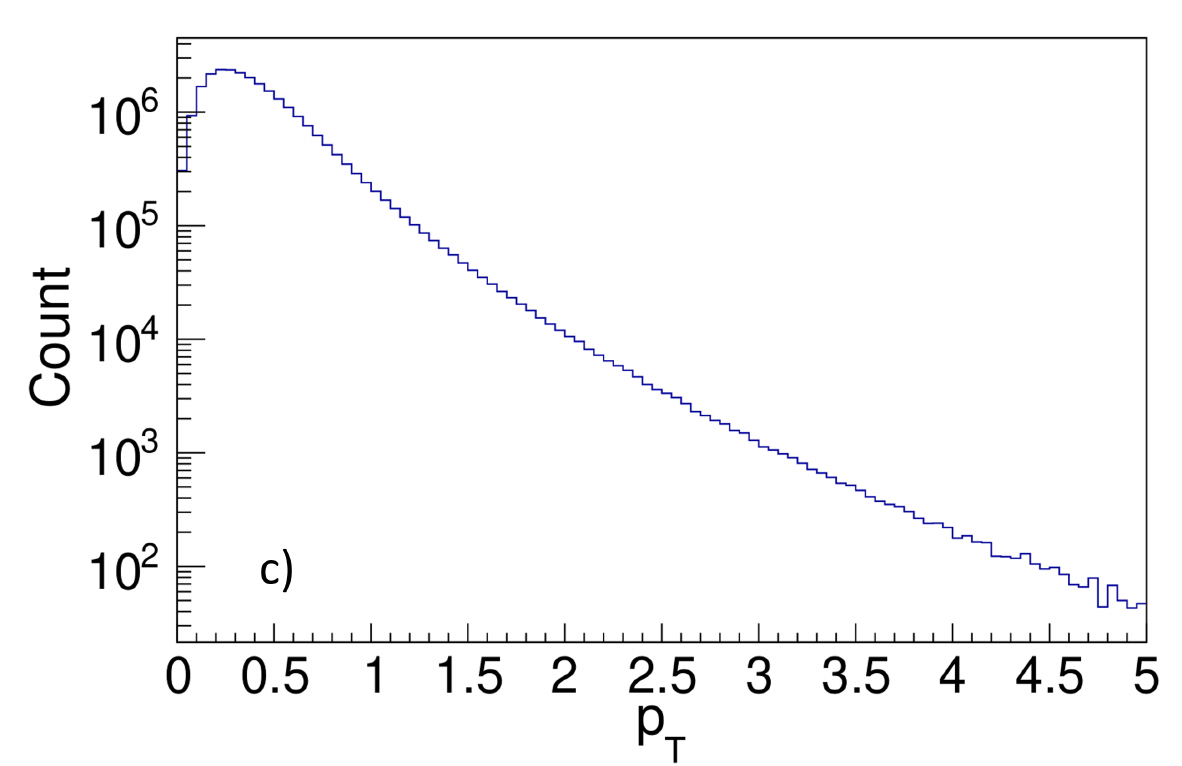}
\caption{ Particle distributions in (a) $\eta$, (b) $\phi$, and (c) $p_T$ for unbiased PYTHIA events. The upper narrow panels 
of the $\eta$ and $\phi$ distributions 
show an amplified view of the 
distributions.}
\label{fig1-1}\end{figure}

\begin{figure}[]\centering
\includegraphics[width = .99\linewidth]{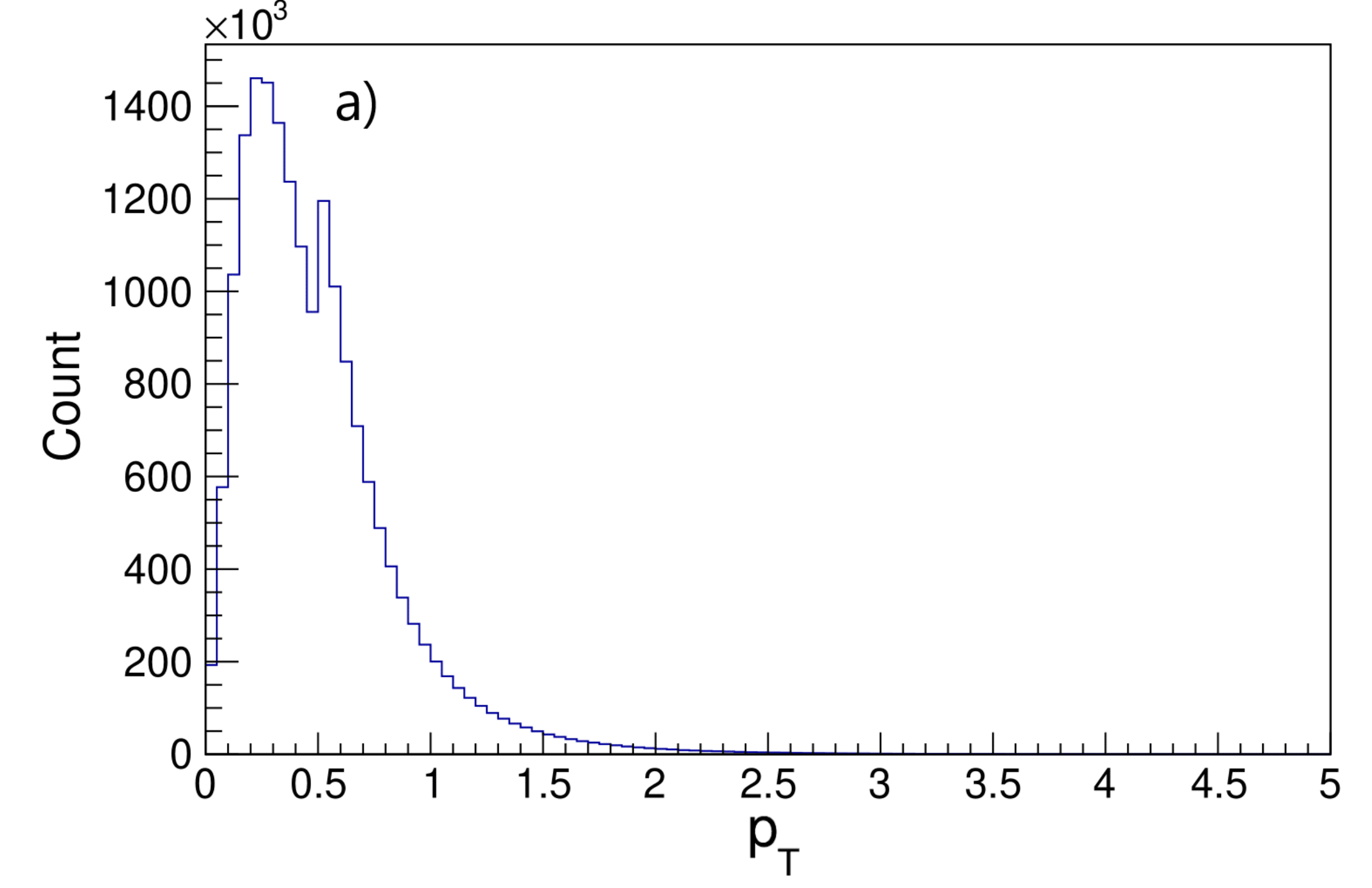}
\includegraphics[width = .99\linewidth]{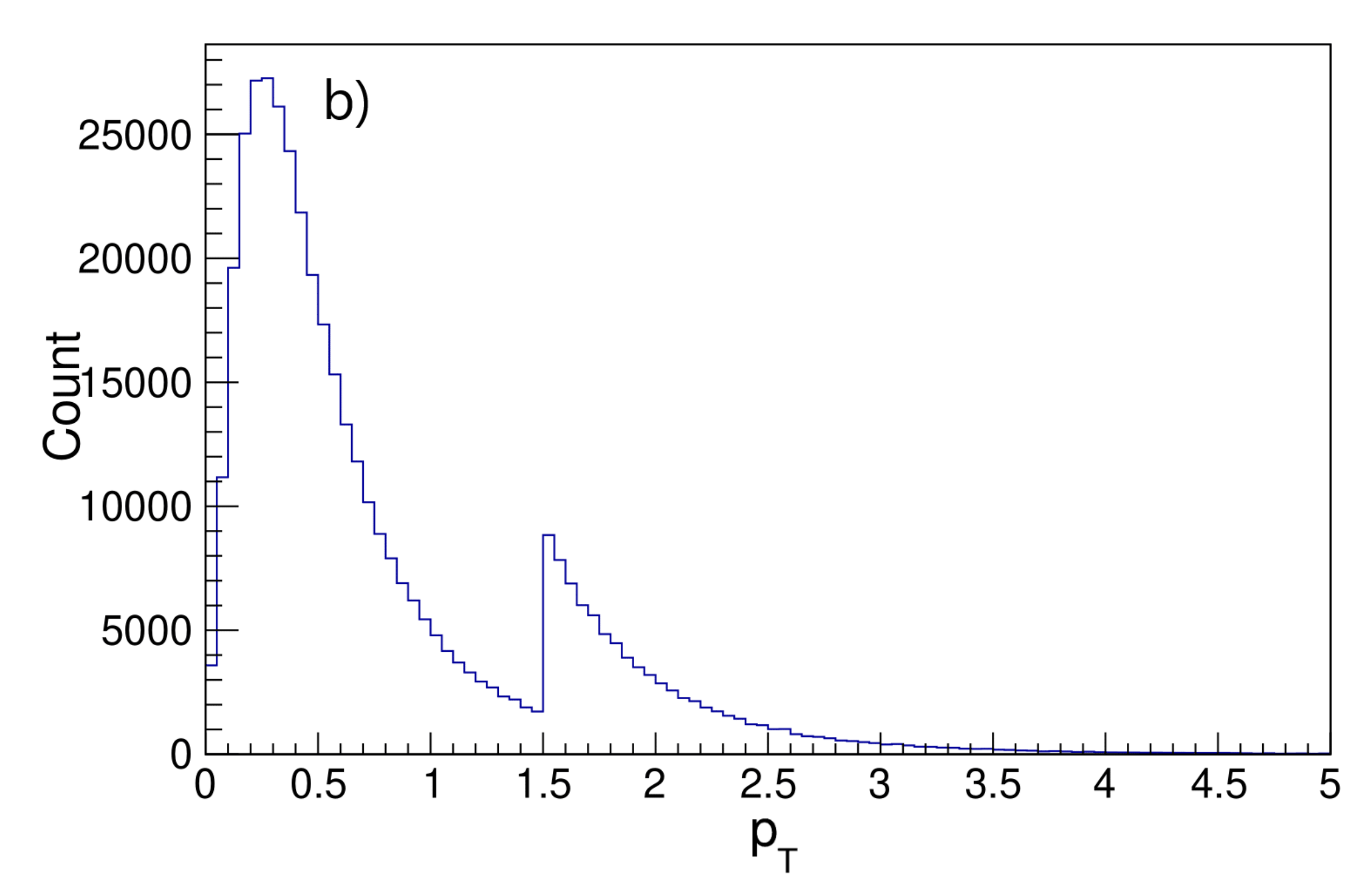}
\caption{Particle  $p_T$ distributions  for PYTHIA events selected with $p_{T0} = 0.5$ (a) and $1.5$ GeV/$c$ (b).}
\label{fig1-2}
\end{figure}

To assess the viability and performance of our $p_T$-seeded clustering algorithm, we conducted evaluations using PYTHIA8.1 simulations for particles produced in $p$+$p$ collisions at $\sqrt{s}=200$ GeV, within the range of $-1 \leq \eta \leq 1$ and $-\pi \leq \phi \leq \pi$. Figure~\ref{fig1-1} depicts the $\eta$, $\phi$, and $p_T$ distributions of the charged particles in the simulation. We have explored three different values of the threshold $p_{T0}$: 0.5, 1, and 1.5 GeV/$c$, and retained events with at least two particles having $p_T \geq p_{T0}$ for correlation analyses. The $p_T$ distributions of all particles in events that passed the $p_{T0}$ cut of 0.5 and 1.5 GeV/$c$  are presented in Fig. \ref{fig1-2}. The discontinuities observed in the $p_T$ distributions at $p_{T0}$ are due to the threshold effect, as events comprising solely low-$p_T$ particles are removed. Despite this, low-$p_T$ particles from the surviving events still contribute to the distribution below $p_{T0}$. Therefore, the distribution above the $p_{T0}$ thresholds maintains its similarity to the original uncut distribution, while the distribution below these thresholds experiences a reduction due to the removal of low-$p_T$ events, thereby causing the observed discontinuity.

For comparison with the PYTHIA data (same events), we construct the mixed events by reassigning $\phi$ and $\eta$ values to particles in each PYTHIA event, while preserving the event's particle multiplicity and each particle's $p_T$ value. The $\phi$ and $\eta$ values are randomly sampled from the original particle's $\phi$ and $\eta$ distributions, shown in Fig. \ref{fig1-1}. This procedure ensures that the resulting particles naturally have the same $p_T$ values as the original particles, the mixed events have the same multiplicities as the original events, and the mixed data set has the same $\phi$ and $\eta$ distribution as the PYTHIA data set. In the following, we study the difference between the same events and the mixed events in terms of two-particle
and two-cluster correlations.

\begin{figure}[]\centering
\includegraphics[width = .99\linewidth]{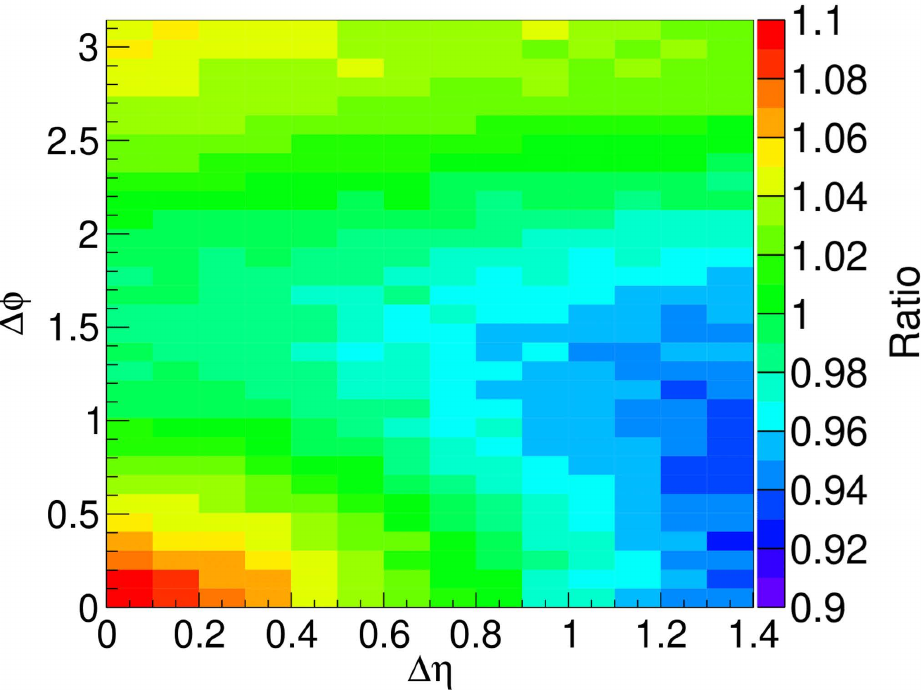}
\includegraphics[width = .99\linewidth]{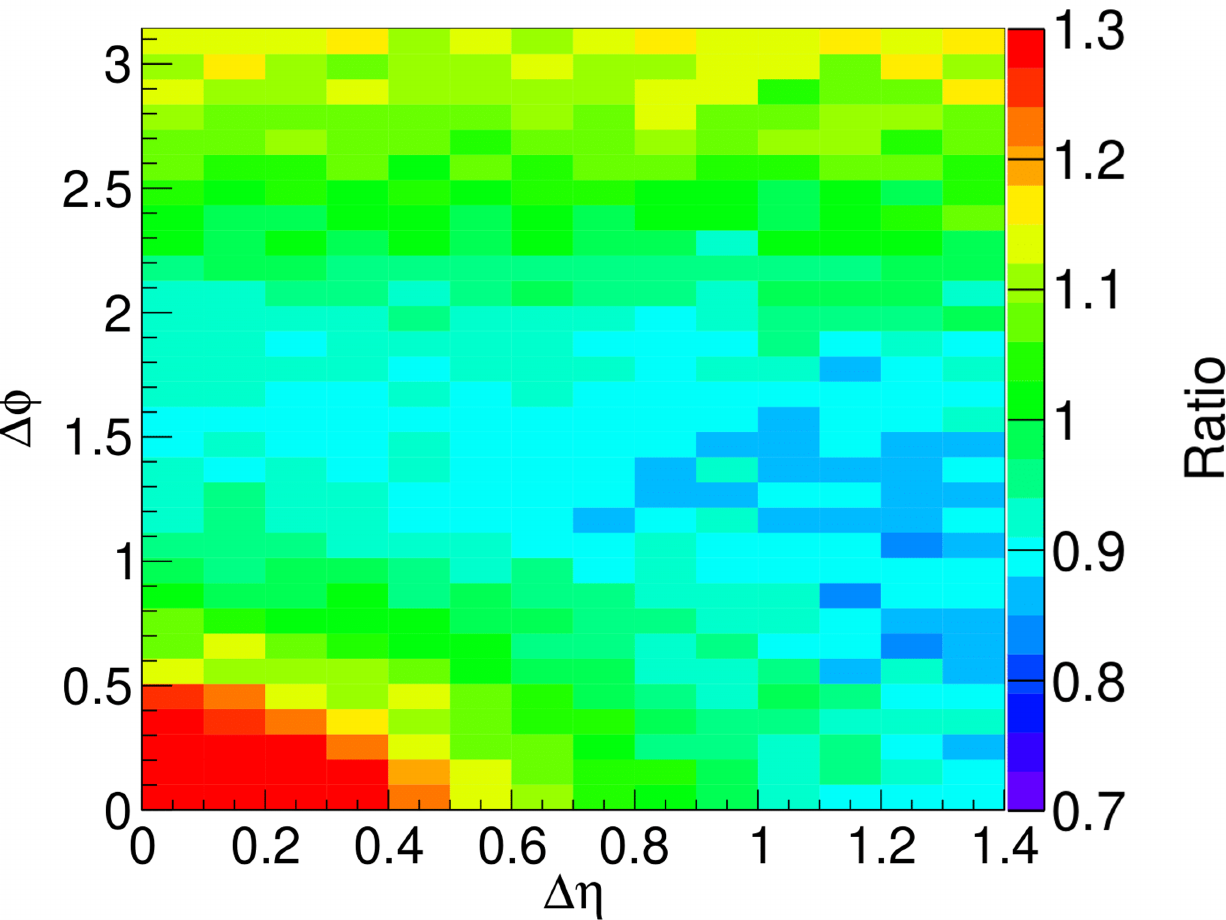}
\caption{Ratio of the two-particle correlation ($\Delta\phi$-$\Delta\eta$) function for the same events to that for the mixed events with $p_{T0} = 0.5$  (upper) and 1.5 GeV/$c$ (lower).
}
\label{fig1-3}\end{figure}

Figure~\ref{fig1-3} shows the ratio of the two-particle correlation ($\Delta\phi$-$\Delta\eta$) function for the same events to that for the mixed events with $p_{T0} = 0.5$  (upper panel) and 1.5 GeV/$c$ (lower panel). Before taking the ratio, the correlations for both the same events and the mixed events have been normalized with the total number of particle pairs. We restrict the analysis to the $\Delta\eta$ range of [0, 1.4] due to the limited number of events that pass the $p_{T0}$ cuts. Although the same events and the mixed events have very similar $\phi$ and $\eta$ distributions, notable differences can be observed between the two data sets in the two-particle correlation. In each panel, the correlation ratio exhibits a near-side peak at $\Delta\eta \approx 0$ and $\Delta\phi \approx 0$, as well as an away-side ridge at $\Delta\phi \approx \pi$, which is consistent with momentum conservation and the analysis of Section III. The away-side ridge becomes more prominent for higher $p_{T0}$, with a magnitude of approximately 1.04 (1.15) for $p_{T0}=$ 0.5 (1.5) GeV/$c$. 

\subsection{Two-Cluster Correlations}

\begin{figure}[]\centering
\includegraphics[width = .99\linewidth]{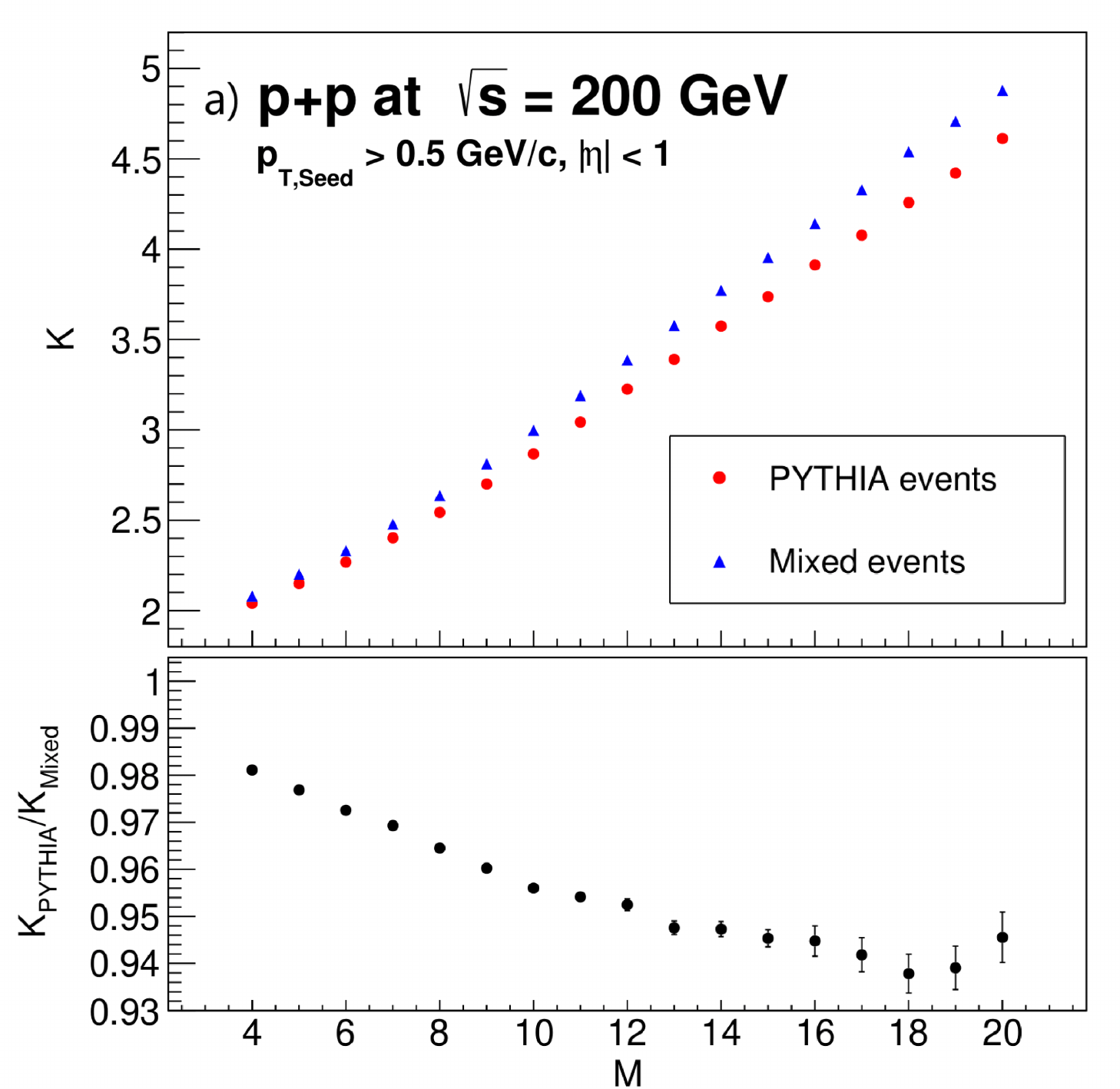}
\includegraphics[width = .99\linewidth]{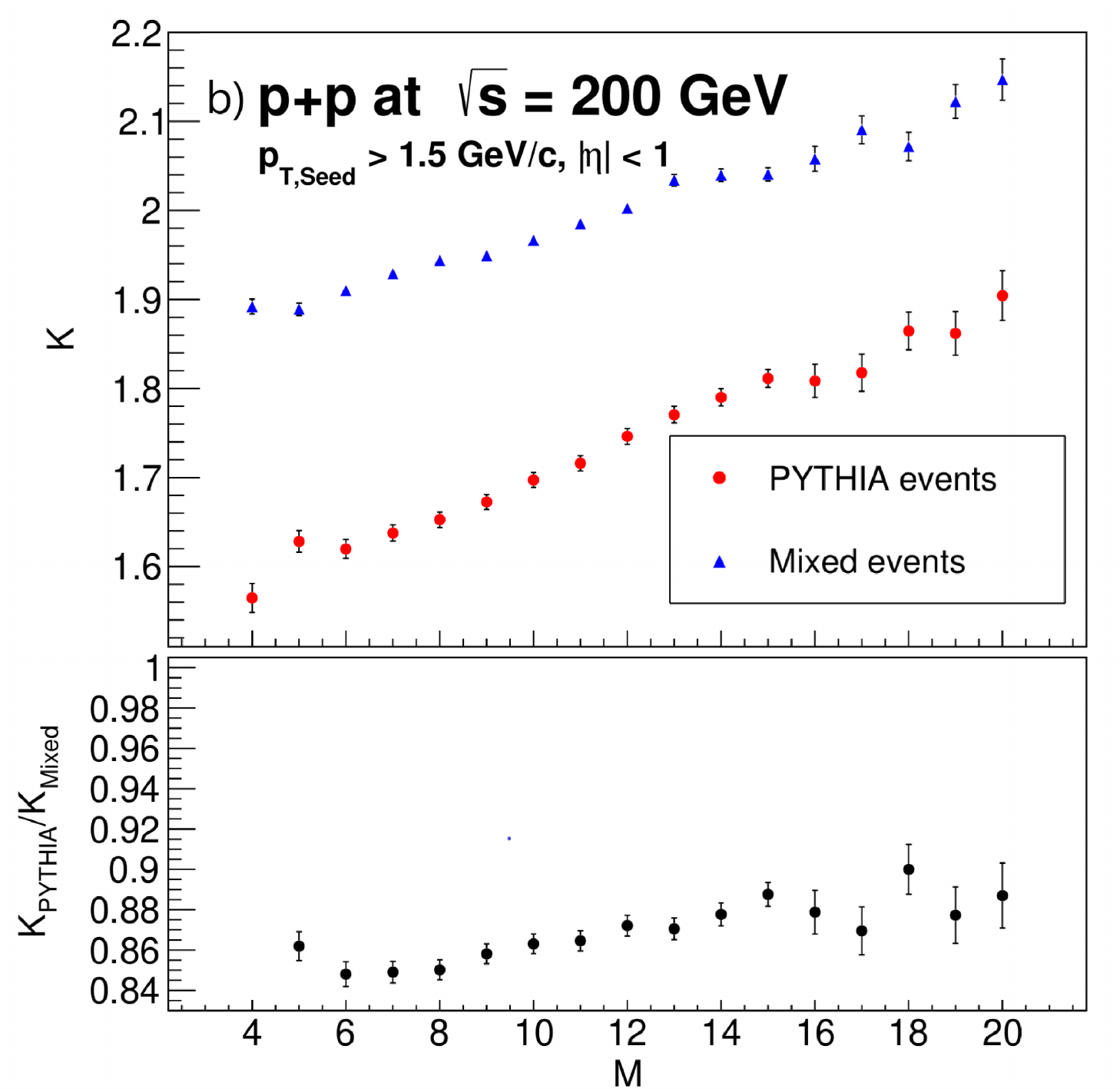}
\caption{Number of clusters, $K$, as a function of multiplicity, $M$, for the same events (red circle) and the mixed events (blue triangle), and the ratio (black circle) of the same events to the mixed events for $p_{T0} =$ (a) 0.5 GeV/$c$  and (b) $1.5$ GeV/$c$.}
\label{fig0-1}\end{figure}

While the two-particle correlation can reveal some interesting properties, the two-cluster correlation is expected to amplify the correlation features related to mini-jets due to the use of the $p_T$-seeded clustering algorithm. Figure~\ref{fig0-1} shows the number of reconstructed clusters, $K$, as a function of multiplicity, $M$, for $p_{T0} =$ (a) 0.5 and (b) 1.5 GeV/$c$. We observe significant differences between the same events and the mixed events, as well as between the cases with different $p_{T0}$ values. When $p_{T0} =$ 0.5 GeV/$c$, the difference between the same events and the mixed events is not prominent at low $M$, and mixed events show a slightly larger $K$ at higher $M$. The ratio, $K_{\rm same}/K_{\rm mixed}$, exhibits a slowly decreasing trend from 0.98 to 0.94. On the other hand, when $p_{T0} =$ 1.5 GeV/$c$, the mixed events yield significantly more clusters than the same events, with $K_{\rm same}/K_{\rm mixed}$ being roughly constant around 0.86. The difference between the same events and the mixed events can be attributed to the clustering property of PYTHIA particles, among which particles belonging to the same mini-jet are likely to be captured by our algorithm to form a cluster. Conversely, particles in the mixed events are uniformly distributed, and clusters are randomly formed. Thus, the algorithm is more effective in clustering mini-jets when using higher $p_{T0}$ values, resulting in a more prominent difference between the same events and the mixed events. $p_{T0} = 0.5$ GeV/$c$ may be insufficient to clearly distinguish the same events and the mixed events, especially at low $M$. 

\begin{figure}[tbhp]\centering
\includegraphics[width = .99\linewidth]{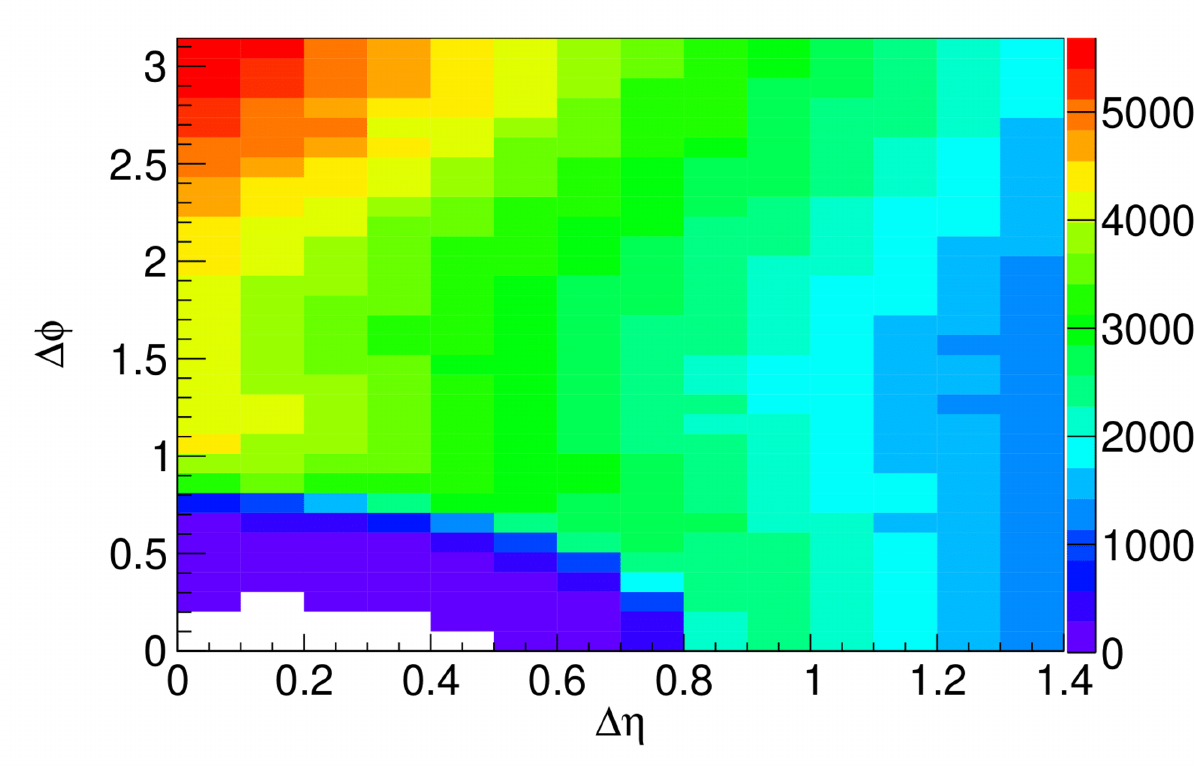}
\includegraphics[width = .99\linewidth]{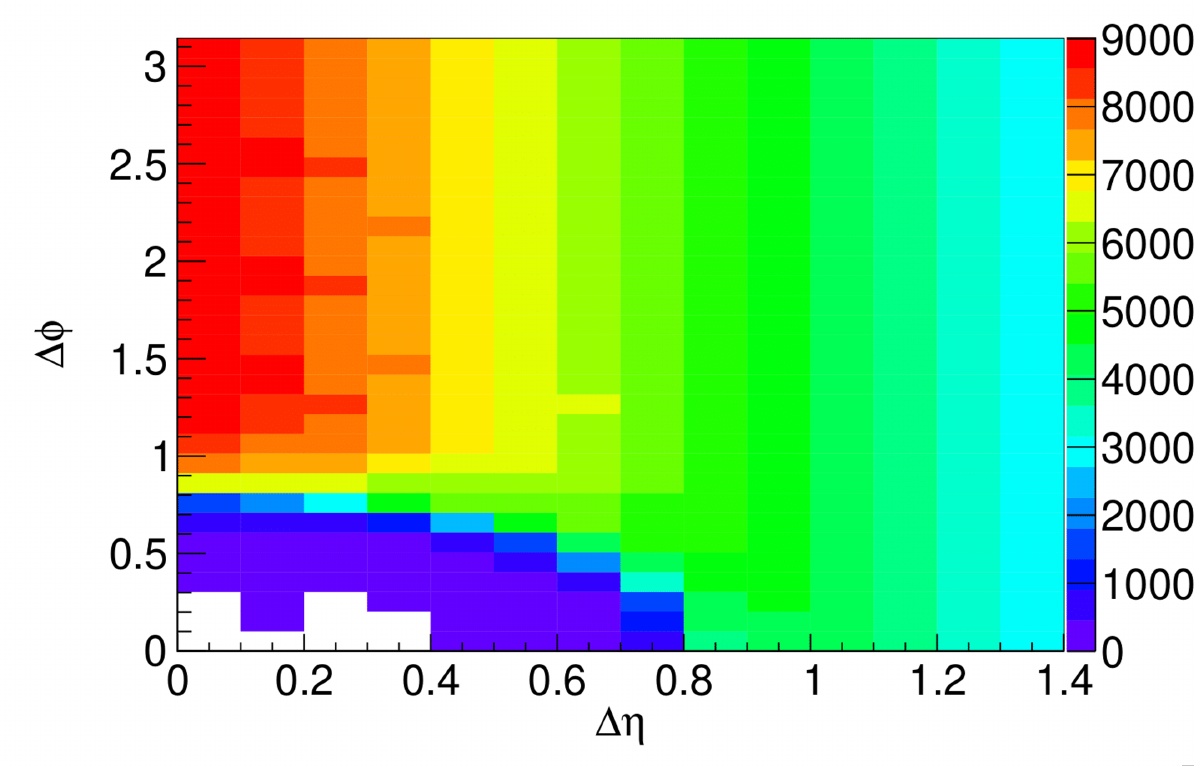}
\caption{Two-cluster correlation ($\Delta\phi$-$\Delta\eta$) for the same events (upper)  and the mixed events (lower) with $p_{T0} = 0.5$ GeV/$c$ and $K = 2$.}
\label{fig2-1}\end{figure}

To investigate the correlation between two clusters, we classify events based on the value of $K$ obtained from the algorithm. Specifically, we analyze the cases of  $K =$ 2, 3, and 4  to demonstrate the dependence on  $K$ and to provide adequate statistical data for further analysis.

\begin{figure}[]\centering
\includegraphics[width = .99\linewidth]{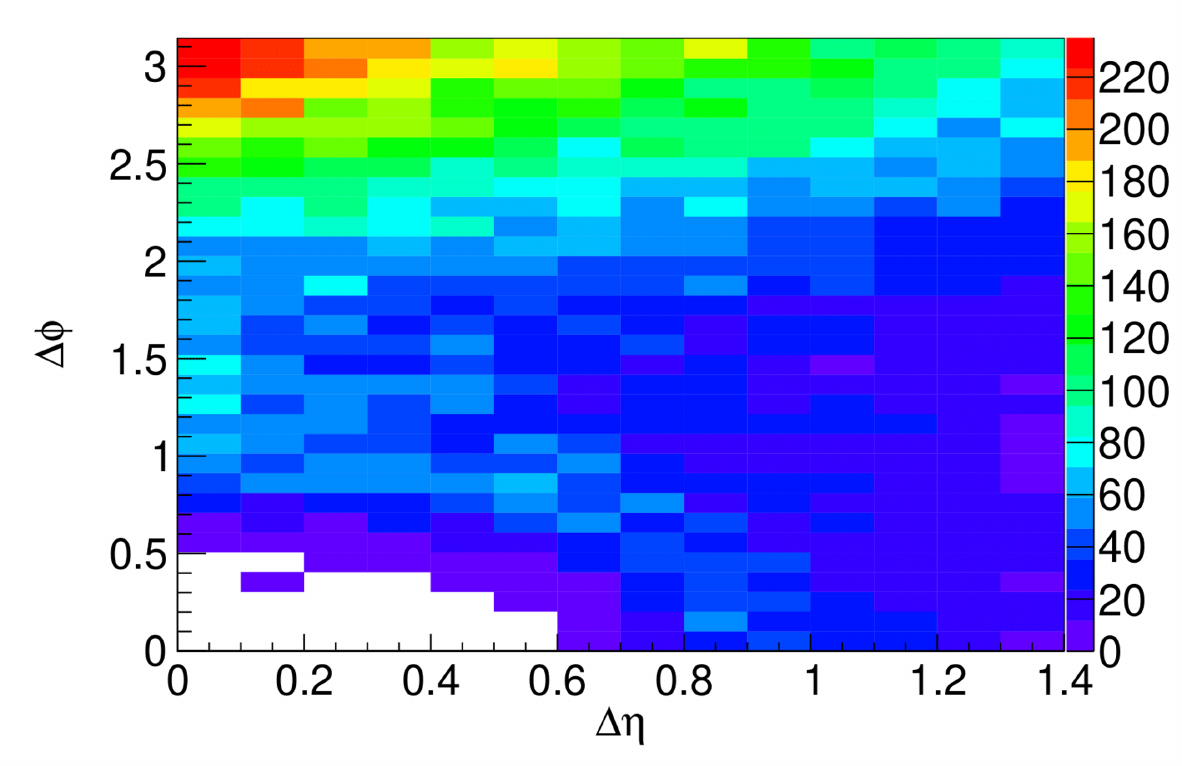}
\includegraphics[width = .99\linewidth]{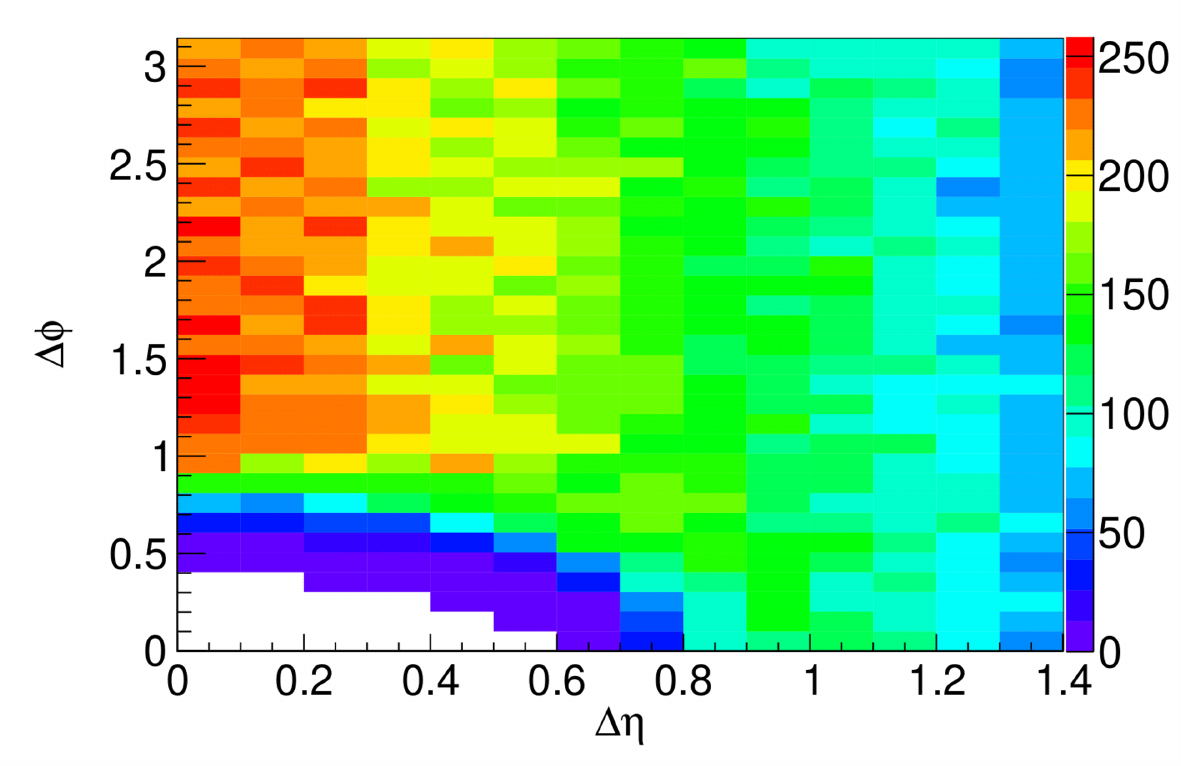}
\caption{Two-cluster correlation ($\Delta\phi$-$\Delta\eta$) for the same events (upper) and the mixed events (lower) with $p_{T0} = 1.5$ GeV/$c$ and $K = 2$.}
\label{fig2-2}\end{figure}

Figures \ref{fig2-1} and \ref{fig2-2} display the two-cluster correlations ($\Delta\phi$-$\Delta\eta$) for the same events (upper) and the mixed events (lower) with $K = 2$ for $p_{T0} = 0.5$ and 1.5 GeV/$c$, respectively. Note that $\Delta\phi$ and $\Delta\eta$ now represent the distances between two cluster centers instead of two particles. The 2D distribution primarily follows the $\Delta\eta$ distribution that approximately takes the shape of a triangle, holds a ridge around $\Delta\eta \approx 0$, and gradually decreases as $\Delta\eta$ increases. The holes around $\Delta\phi \approx 0$ and $\Delta\eta \approx 0$ in these distributions result from the exclusivity of the clusters since two clusters tend to merge in our clustering algorithm if their distance is small. The same-event data exhibit an away-side ridge at $\Delta\phi \approx \pi$, which is absent in the mixed-event data. Specifically, at a fixed value of $\Delta\eta$, the number of entries around $\Delta\phi \approx \pi$ is significantly greater than that around $\Delta\phi \approx \pi/2$ in the same events, whereas it remains nearly uniform along $\Delta\phi$ in the mixed events.  The threshold $p_{T0}$ also impacts these distributions. A significantly higher away-side peak appears in the same events for $p_{T0} = 1.5$ GeV/$c$ (in Fig.~\ref{fig2-2}) than for $p_{T0} = 0.5$ GeV/$c$ (in Fig.~\ref{fig2-1}).

\begin{figure}[]\centering
\includegraphics[width = .99\linewidth]{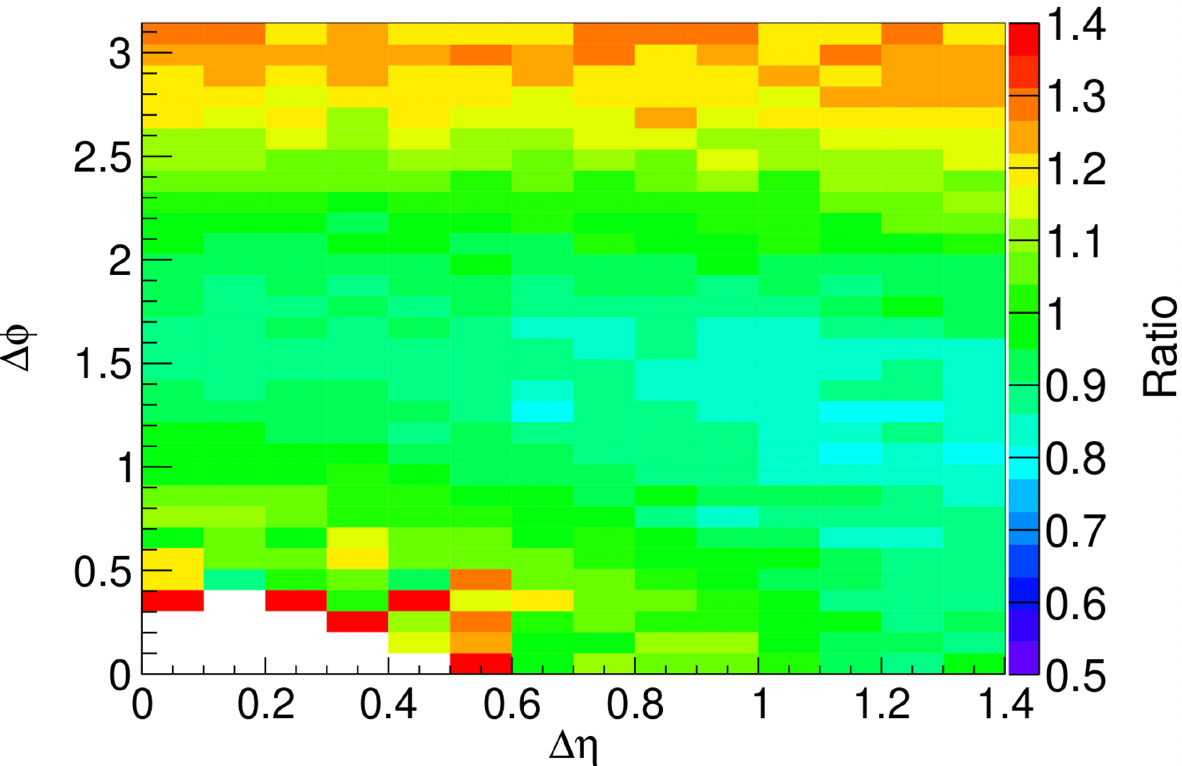}
\includegraphics[width = .99\linewidth]{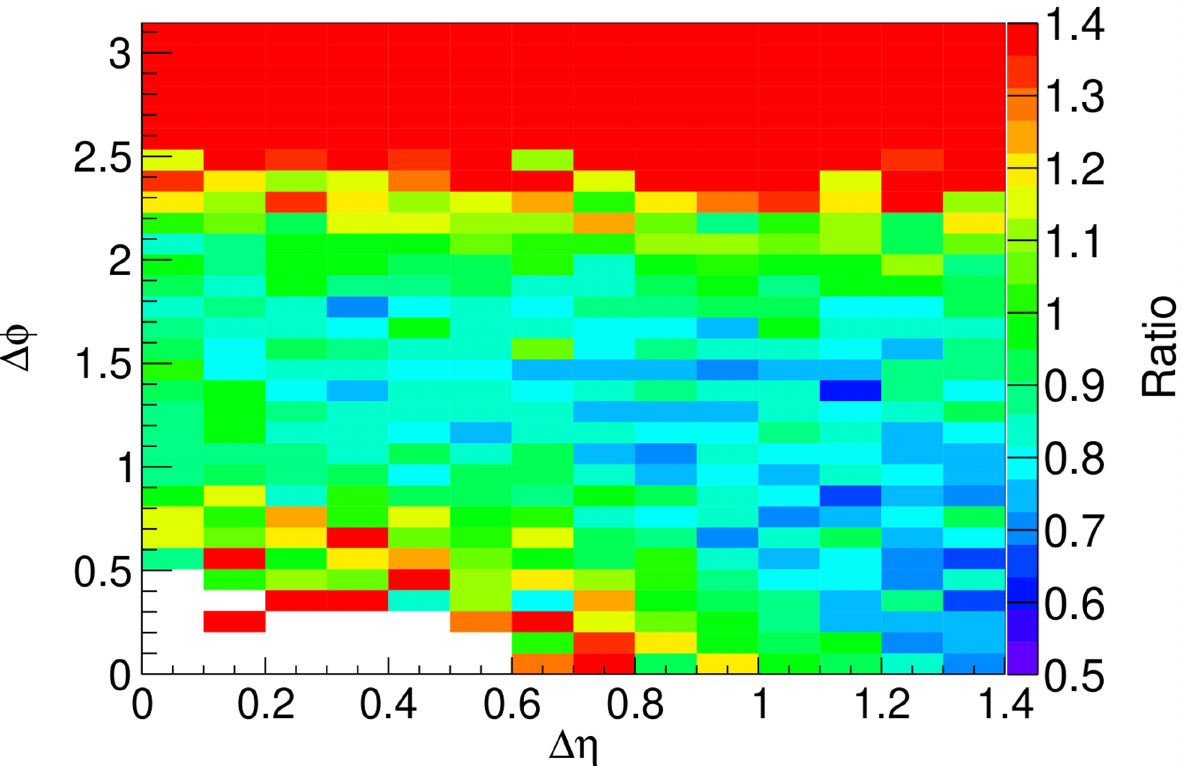}
\caption{Ratio of the two-cluster correlation ($\Delta\phi$-$\Delta\eta$) function for the same events to that for the mixed events for events selected with $K = 2$ for $p_{T0} = 0.5$ (upper) and 1.5 GeV/$c$ (lower).}
\label{fig2-3}\end{figure}

Figure~\ref{fig2-3} shows the ratio of the two-cluster correlation for the same events to that for the mixed events with $K = 2$ for $p_{T0} = 0.5$ (upper) and $1.5$ GeV/$c$ (lower). Before taking the ratio, we have normalized the 2D correlations for the same events and the mixed events with their corresponding total numbers of cluster pairs, respectively,  similar to the approach used for the ratio in the two-particle correlations in Fig.~\ref{fig1-3}. Except for the presence of holes at $\Delta\phi \approx 0$ and $\Delta\eta \approx 0$, the two-cluster correlations qualitatively resemble the two-particle correlations in Fig.~\ref{fig1-3} and reveal more prominent away-side ridges around $\Delta\phi \approx \pi$.  Remarkably, the away-side ridge appears consistently and uniformly for the two-cluster correlations, even for $p_{T0} = 0.5$ GeV/$c$, and is enhanced for the higher $p_{T0} = 1.5$ GeV/$c$. The magnitude of the ratio within the away-side range is about 1.25 (1.4) for $p_{T0} = 0.5$ (1.5) GeV/$c$.

\begin{figure}[]\centering
\includegraphics[width = .99\linewidth]{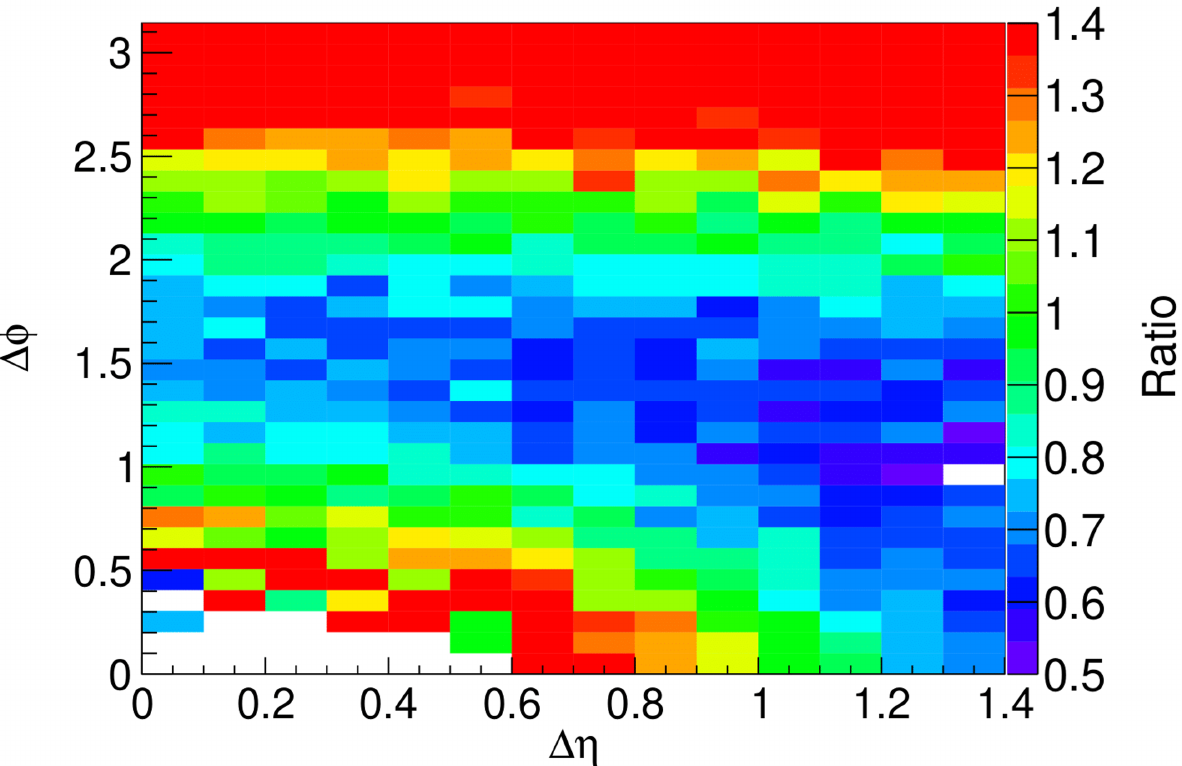}
\includegraphics[width = .99\linewidth]{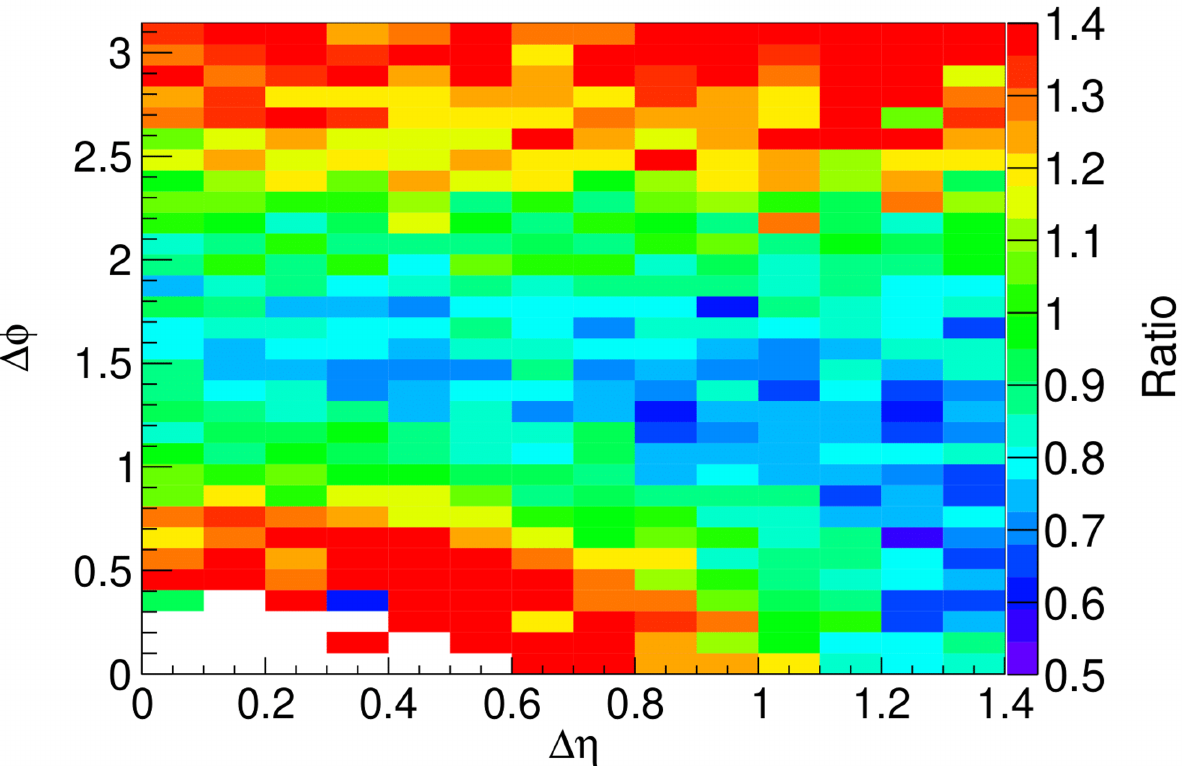}
\includegraphics[width = .99\linewidth]{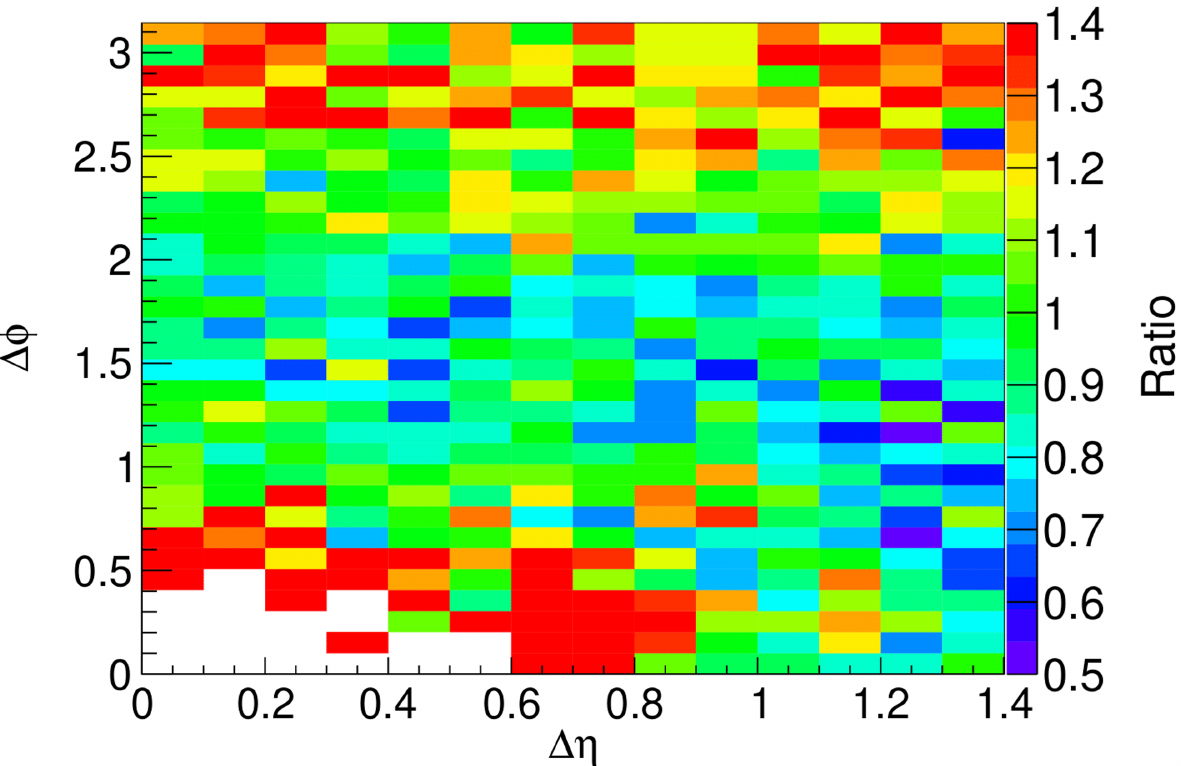}
\caption{Ratio of the two-cluster correlation ($\Delta\phi$-$\Delta\eta$) function for the same events to that for the mixed events with $p_{T0} = 1$ GeV/$c$, and $K =$  2 (upper), 3 (middle), and 4 (lower).}
\label{fig2-4}\end{figure}

Figure~\ref{fig2-4} illustrates the $K$ dependence of the ratio of 
the two-cluster correlation ($\Delta \phi-\Delta \eta$) function 
for the 
same events to that 
for the
mixed events.
with $p_{T0} = 1$ GeV/$c$. The upper, middle, and lower panels correspond to $K$ values of 2, 3, and 4, respectively. As $K$ increases, the magnitude of the away-side ridge in the ratio decreases. 
The away-side ridge is likely due to 
momentum conservation, which is diluted by increasing $K$.

\begin{figure*}[]\centering
\subcaptionbox{Two-particle correlation}{%
\includegraphics[width = .33\linewidth]{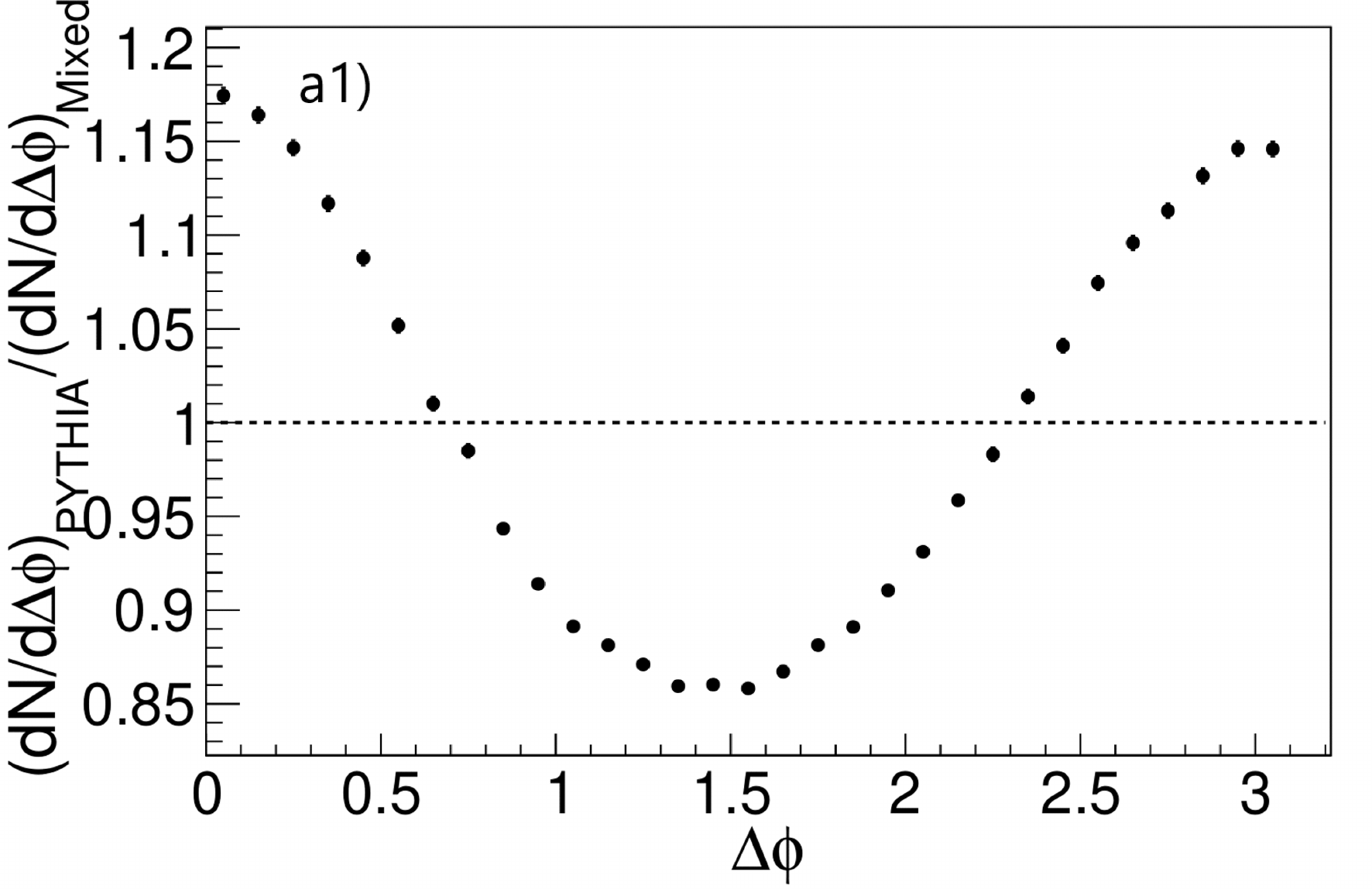}
\includegraphics[width = .33\linewidth]{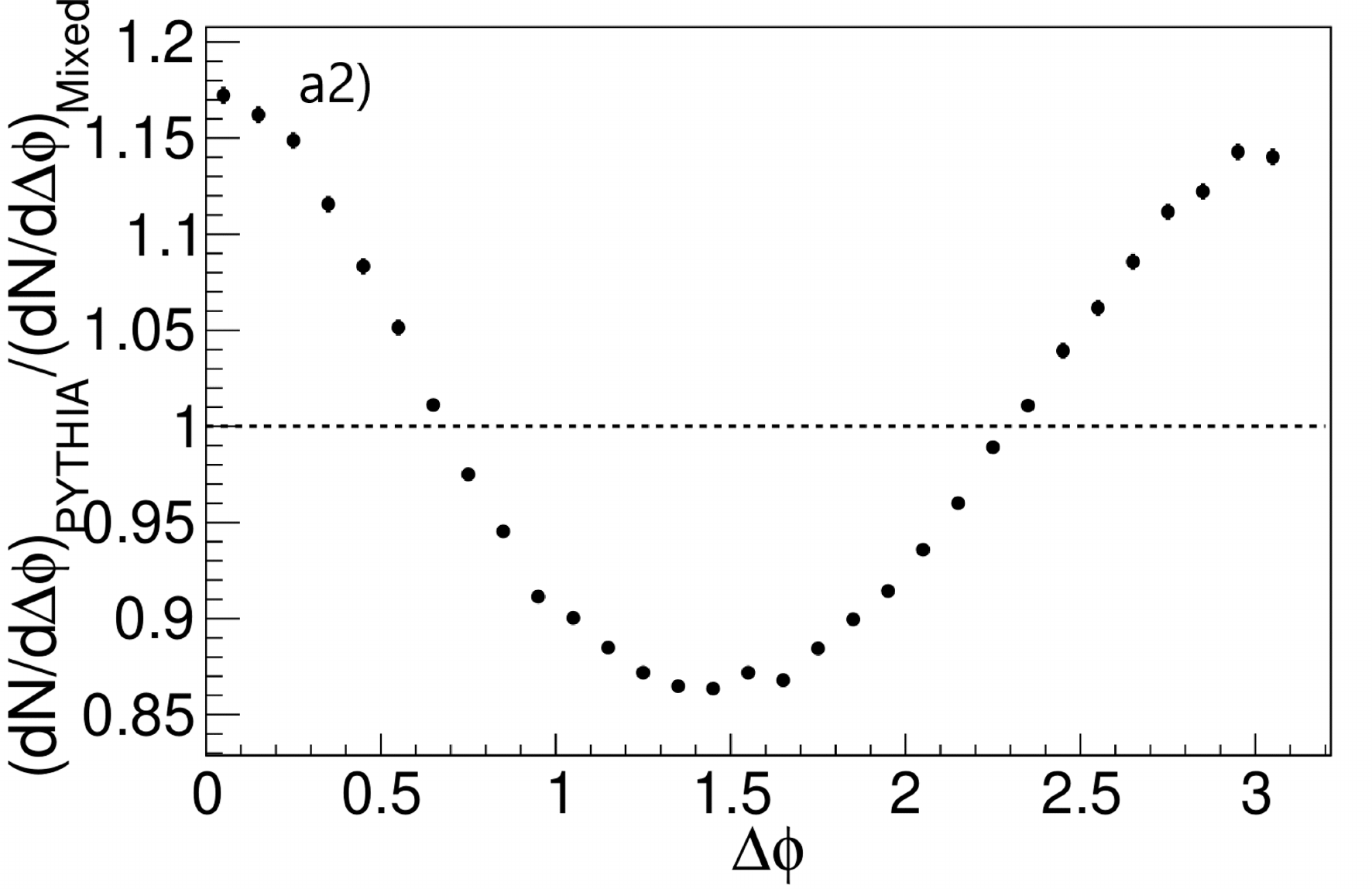}
\includegraphics[width = .33\linewidth]{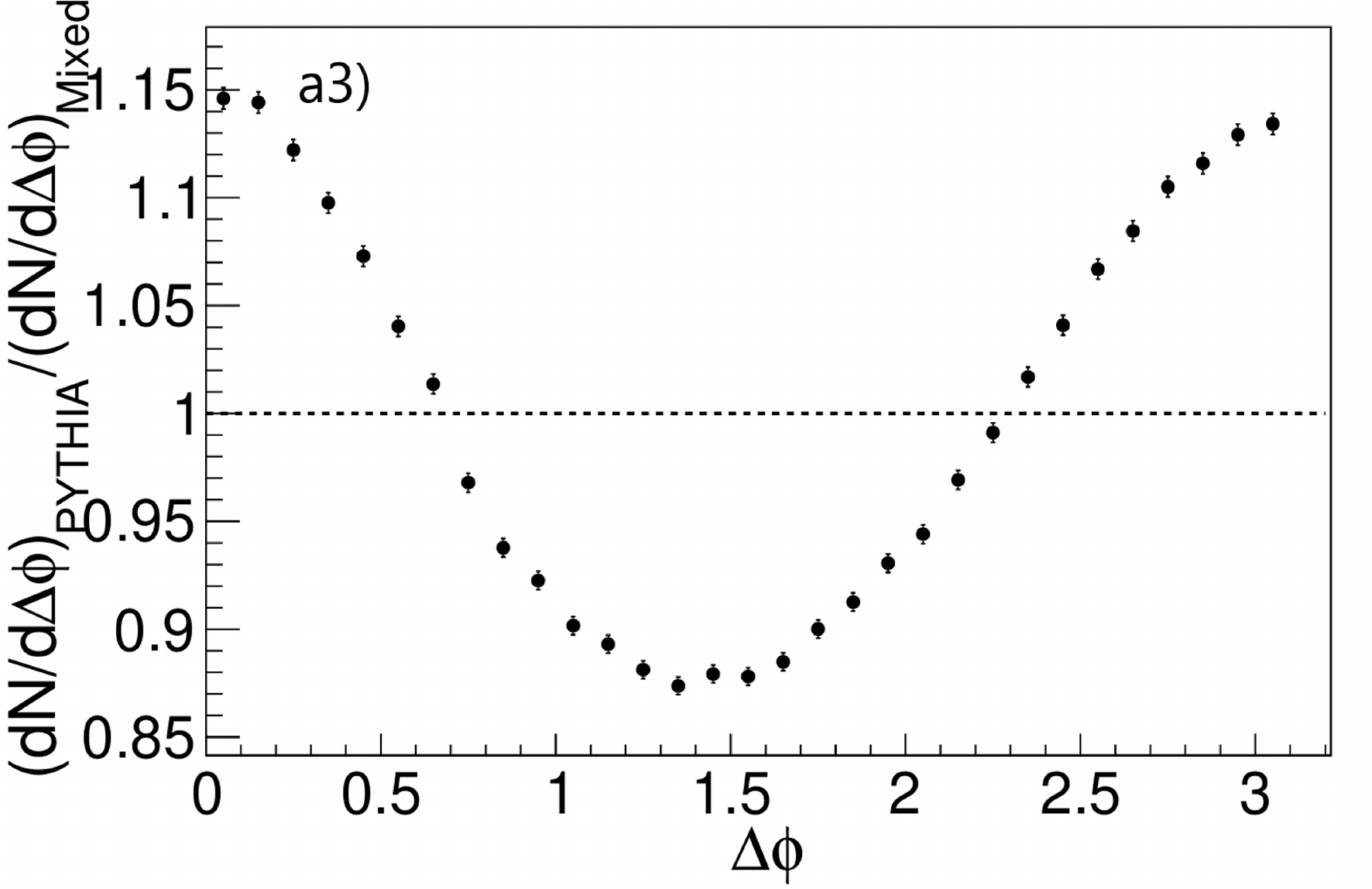}}
\subcaptionbox{Two-cluster correlation}{%
\includegraphics[width = .33\linewidth]{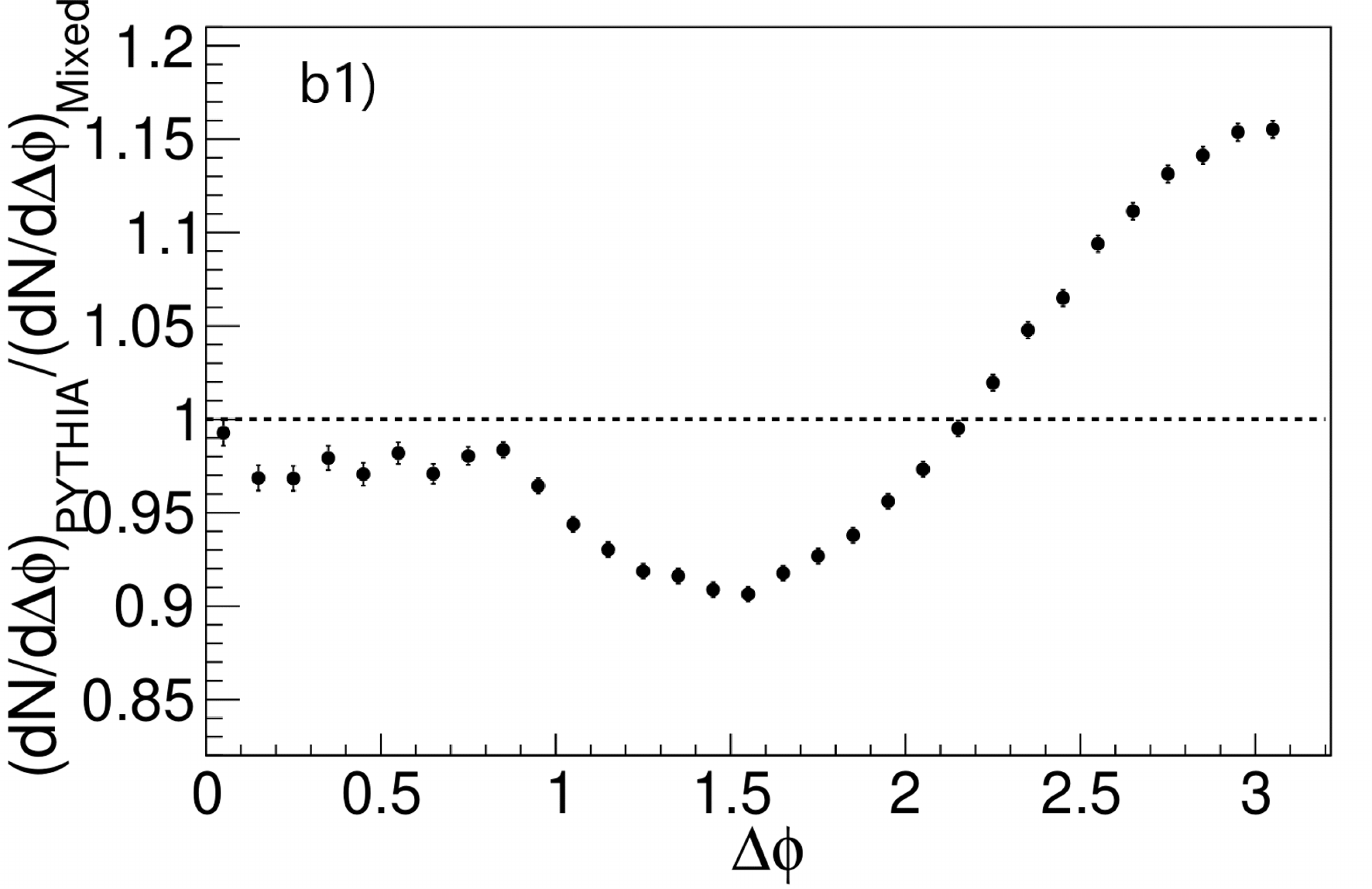}
\includegraphics[width = .33\linewidth]{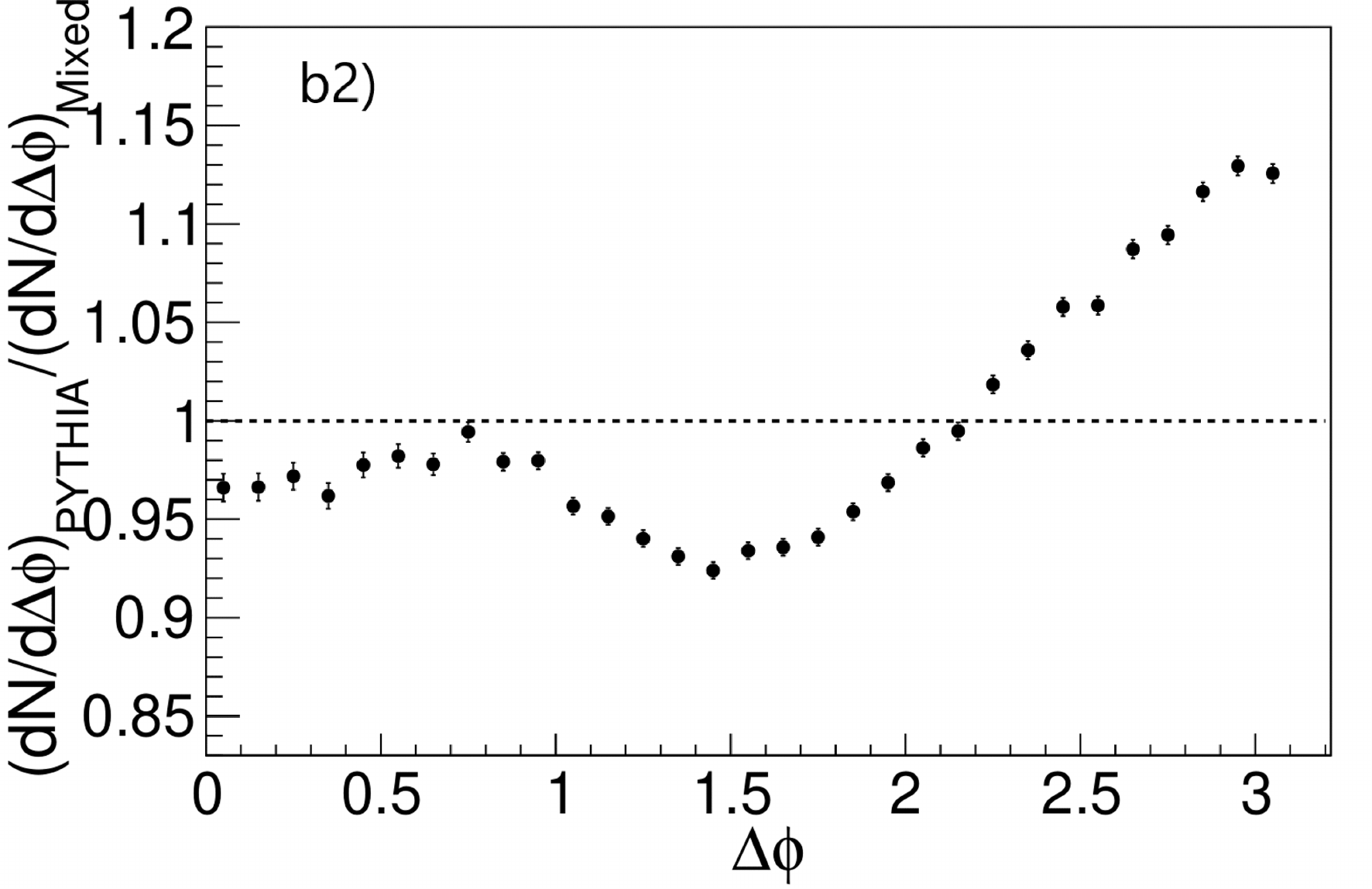}
\includegraphics[width = .33\linewidth]{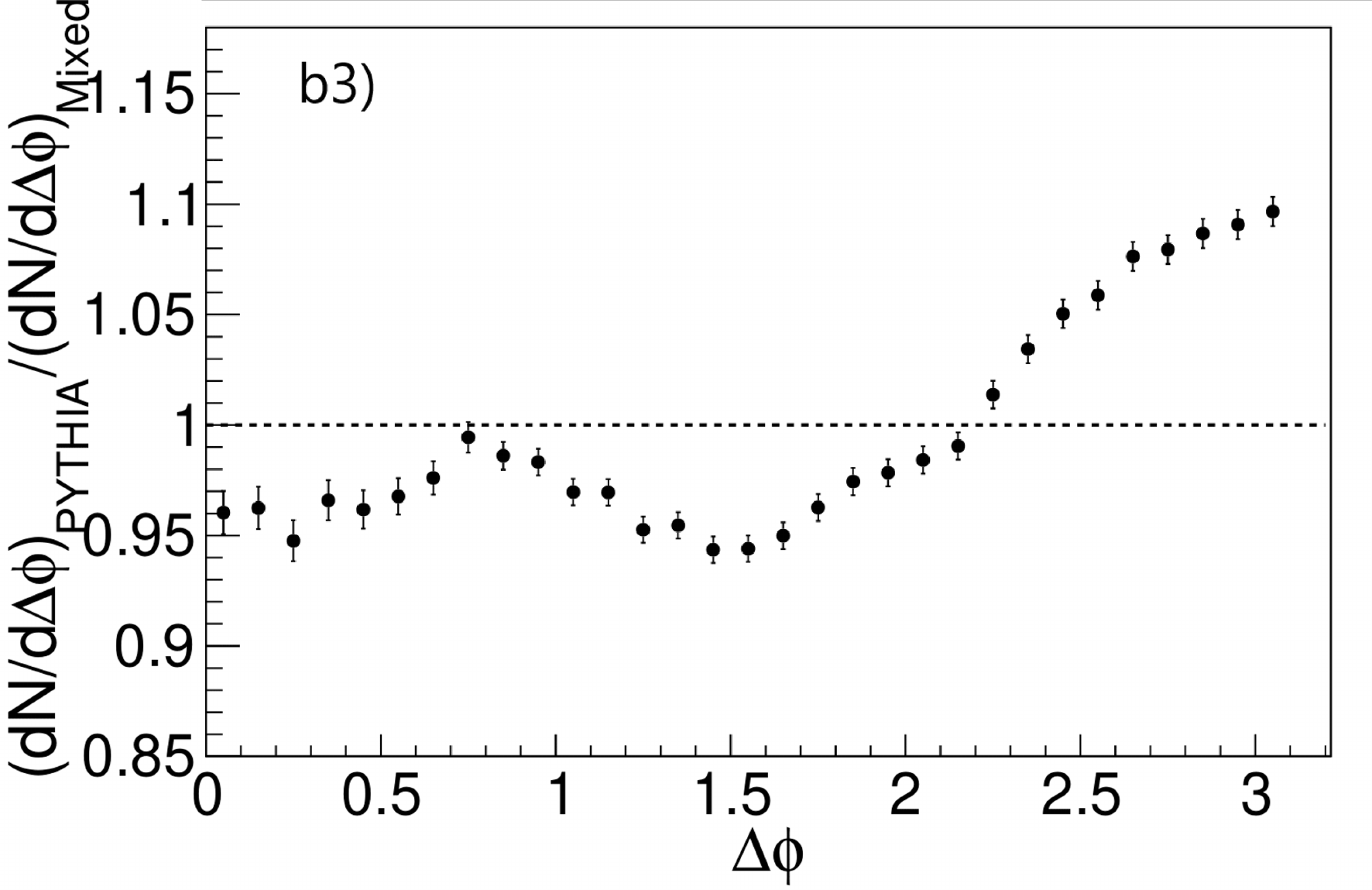}}
\caption{Ratio of the (a) two-particle or (b) two-cluster correlation ($\Delta\phi$) function for the same events to that for the mixed events with $p_{T0}=$ 0.5 GeV/$c$ for the multiplicity ranges of [5, 8] (left), [9, 12] (middle) and [13, 16] (right).}
\label{fig2-5}\end{figure*}

\begin{figure*}[]\centering
\subcaptionbox{Two-particle correlation}{%
\includegraphics[width = .33\linewidth]{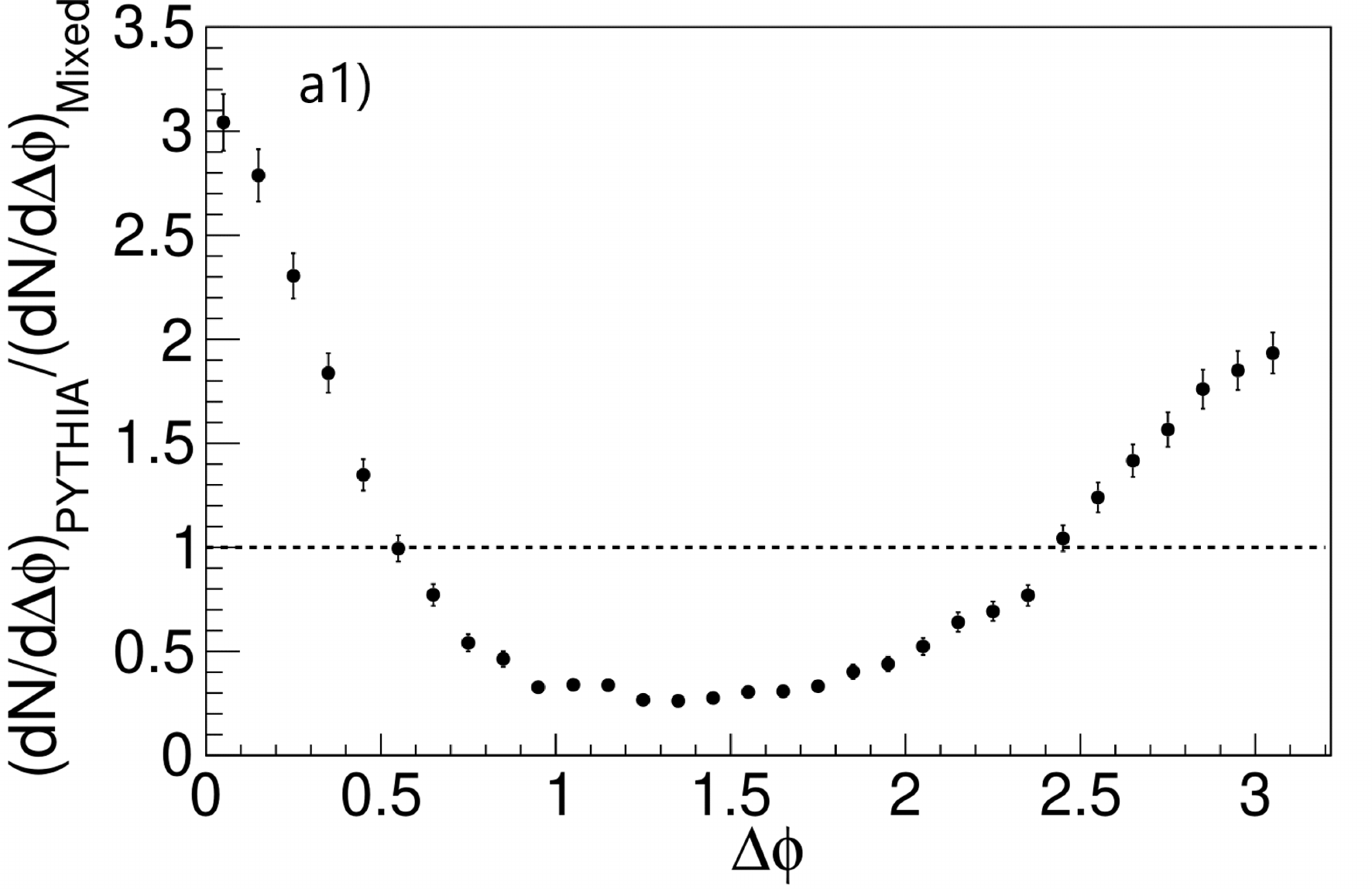}
\includegraphics[width = .33\linewidth]{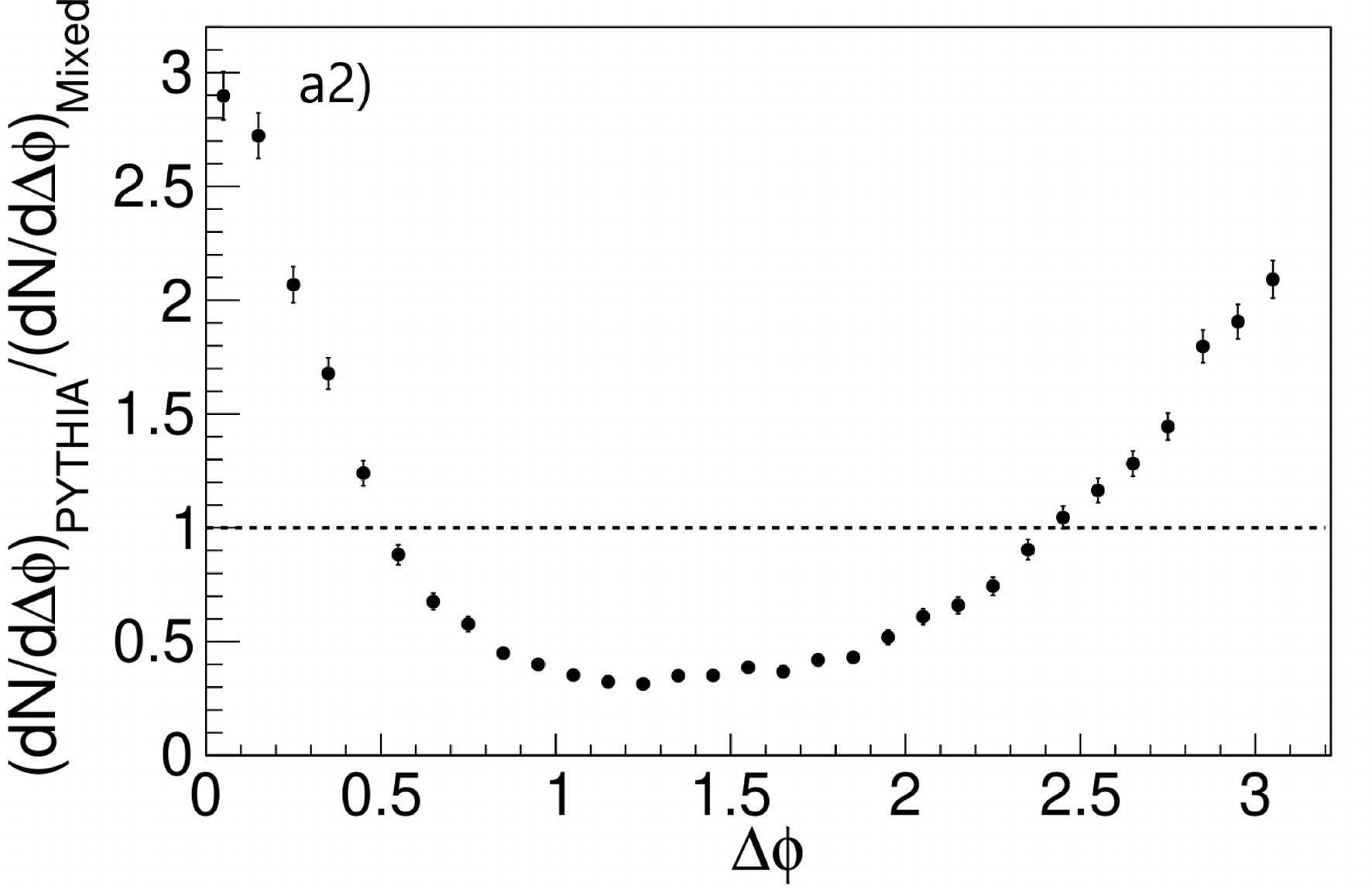}
\includegraphics[width = .33\linewidth]{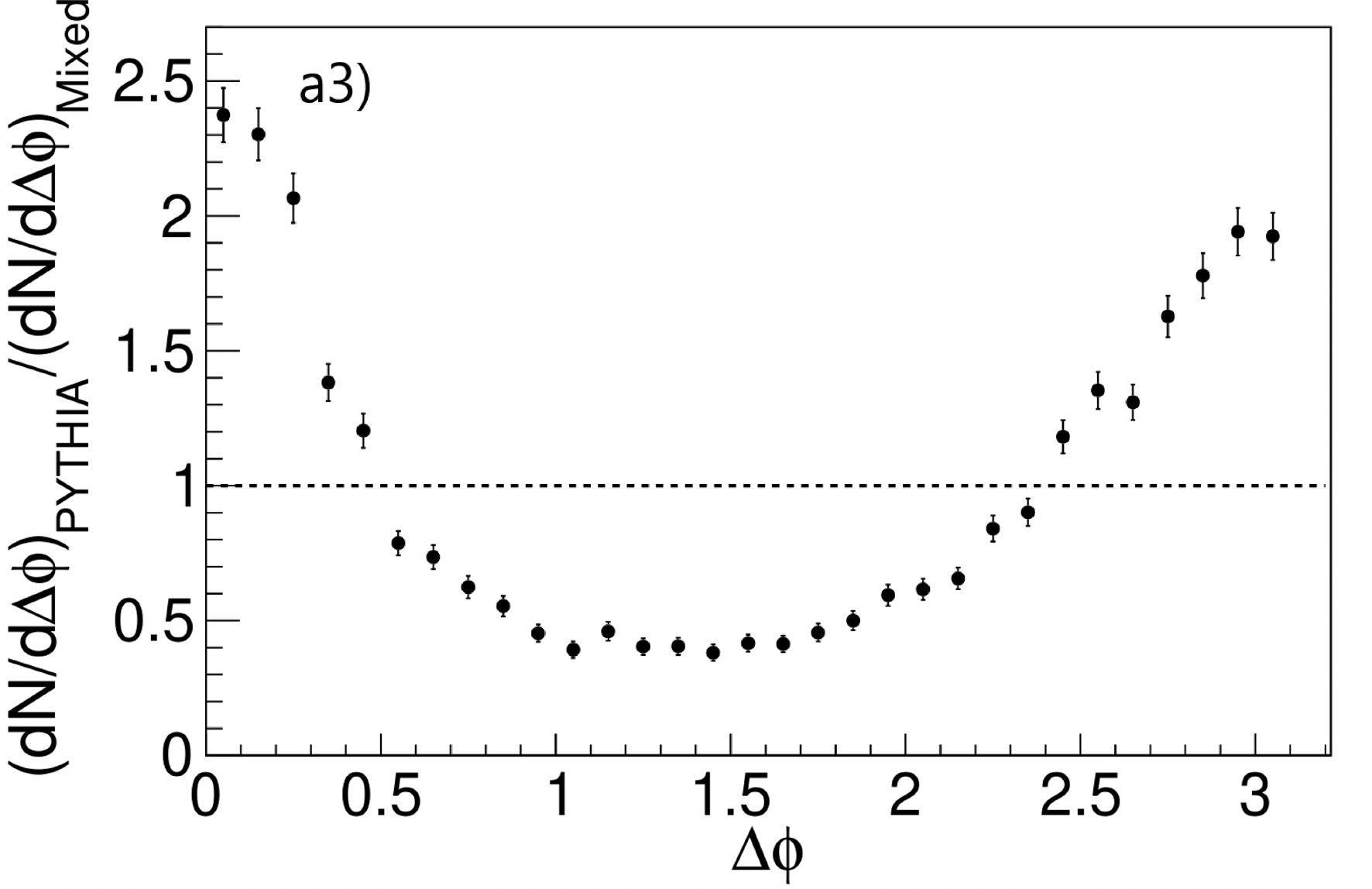}}
\subcaptionbox{Two-cluster correlation}{%
\includegraphics[width = .33\linewidth]{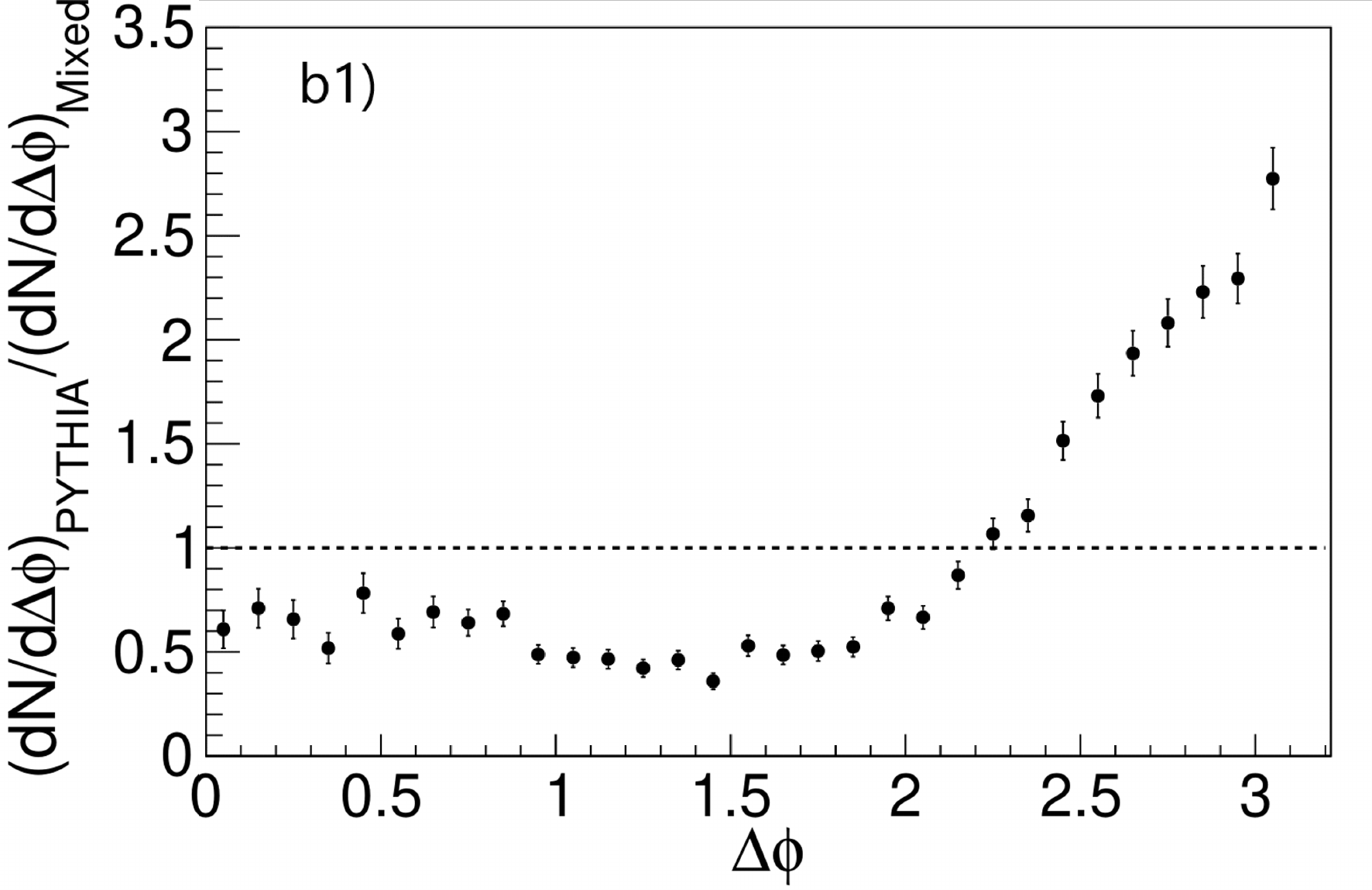}
\includegraphics[width = .33\linewidth]{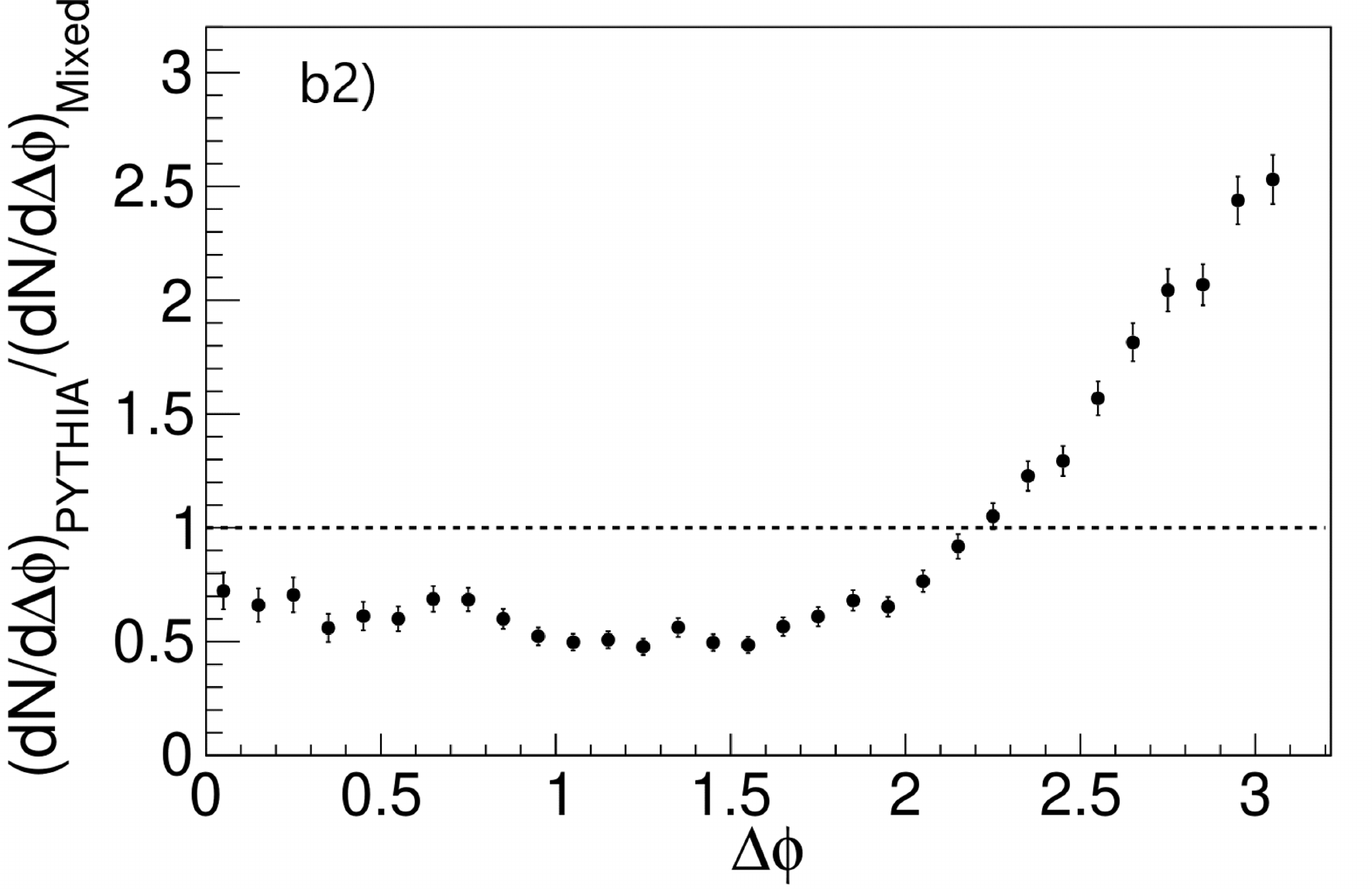}
\includegraphics[width = .33\linewidth]{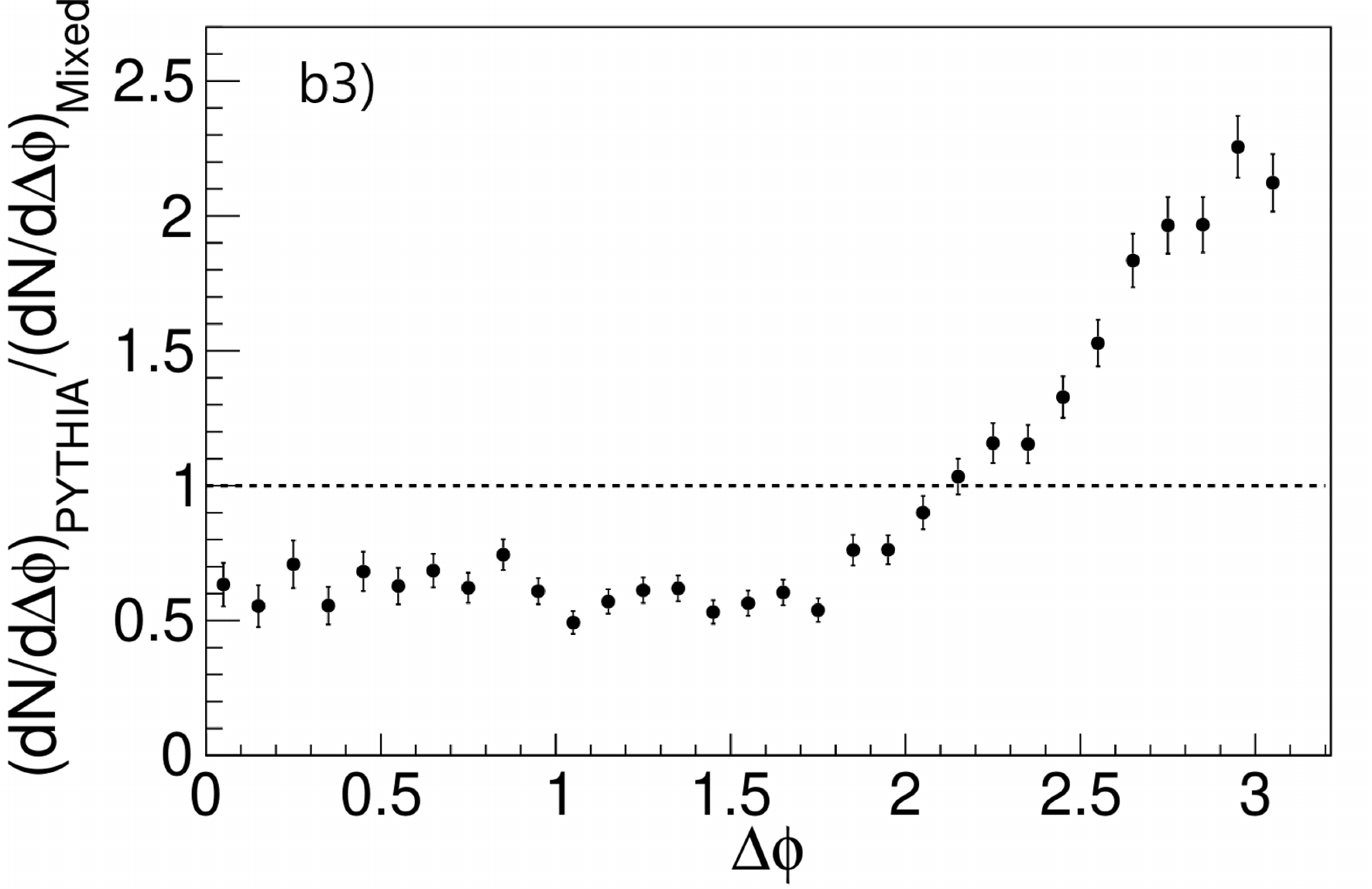}}
\caption{Same as Fig.~\ref{fig2-5}, except for $p_{T0}=$ 1.5 GeV/$c$.}
\label{fig2-6}\end{figure*}

The effectiveness of the $p_{T}$-seeded clustering algorithm can be evaluated by comparing the two-particle and two-cluster correlations, particularly by examining the away-side ridge, and reducing the correlation function to 1D, which simplifies the assessment of the correlation strength. To eliminate bias, both types of correlations are calculated using events with similar multiplicities. Figures~\ref{fig2-5} and \ref{fig2-6} present the ratio of the $\Delta\phi$ distribution for the same events to that for the mixed events with $p_{T0}=$ 0.5 and 1.5 GeV/$c$, respectively, for the multiplicity ranges of [5, 8], [9, 12] and [13, 16]. The upper panels display the two-particle correlations, and the lower ones show the two-cluster correlations. The two-particle correlation shows a near-side peak, which is not present in the two-cluster correlation due to the exclusivity of clusters, as shown by the holes in the 2D correlations in Figs.~\ref{fig2-1}--\ref{fig2-4}. For $p_{T0} =$ 0.5 GeV/$c$, the away-side peak decreases with increased multiplicity, and the peaks in the two-cluster correlations are lower than those in the corresponding two-particle correlations. However, for $p_{T0} =$ 1.5 GeV/$c$, the two-cluster correlations consistently show higher away-side peaks than the two-particle correlations. Hence, by selecting an appropriate $p_{T0}$, the clustering algorithm is more effective in revealing the mini-jet properties compared with the two-particle correlations.

\section{ Difference with the anti-$k_T$ algorithm}
\label{Sec:VI}

The anti-$k_T$ algorithm~\cite{antikt}  is a sequential recombination approach, based on the distance measure,
\begin{equation}
d_{ij} = \min(p^{-2}_{Ti},p^{-2}_{Tj}) \times \frac{R^2_{ij}}{R^2},
    \label{func1}
\end{equation}
in combination with a boundary condition: 
\begin{equation}
    d_{iB} = p^{-2}_{Ti},
    \label{func2}
\end{equation}
where $R^2_{ij} = (\eta_i-\eta_j)^2+(\phi_i-\phi_j)^2$. The first component   $d_{ij}$ of Eq.~(\ref{func1}) serves as a filter that selects particles suitable for inclusion in a cluster. The use of a power of $-2$ for $p_T$ prioritizes the clustering of high-$p_T$ particles. The second component of Eq. (\ref{func2}) sets the criterion for either initiating or halting the recombination process. To satisfy the recombination condition $d_{ij}<d_{iB}$,  the value of $R_{ij}$ must be smaller than the cluster radius $R$. If this condition is not satisfied, the recombination process is terminated. The algorithm iterates through the recombination process until only a few objects remain, and these remaining objects are then referred to as jets. 

While not strictly a cone algorithm, the anti-$k_T$ algorithm exploits the existence of high-$p_T$ particles to stabilize the shapes of the resulting clusters, producing outcomes that resemble a cone-shaped pattern. This is especially evident under typical circumstances where the high-$p_T$ particles exhibit substantial differences from soft particles. Due to the considerable contrast in the $p_T$ values between hard and soft particles, the clustering process is relatively insensitive to the inclusion of soft particles in the jet. However, the anti-$k_T$ algorithm may not be ideally suited for the mini-jet condition, which features considerable presence of soft particles and the indistinct demarcation between high- and low-$p_T$ particles. The presence of non-jet soft particles makes it hard to clearly define mini-jets when we use the anti-$k_T$ algorithm. In contrast, the $p_T$-seeded clustering method is a cone algorithm that allows the distinct separation of the mini-jet with other soft particles using high-$p_T$ seeds, providing advantages in the low $p_T$ regime.

The $p_T$-seeded clustering method is designed to exclude the influence of soft particles in the mini-jet environment. The effect of this method can be observed in cases with $p_{T0}$ values of 0.5 GeV/$c$ and 1.5 GeV/$c$. In particular, clusters with $p_{T0} =$ 0.5 GeV/$c$ display minimal discrepancies in the two-cluster correlation between the same events and the mixed events (in Figure ~\ref{fig2-3}). However, increasing $p_{T0}$ to 1.5 GeV/$c$ reveals noticeable differences in the correlation functions, emphasizing the influence of soft particles in the mini-jet environment and demonstrating the enhancement provided by the $p_T$-seeded clustering algorithm. The $p_T$-seeded clustering method includes all particles in the vicinity of the seeded cluster center, effectuating a static clustering approach that accurately captures the original particle configuration. Unlike the dynamic, continual particle-merging process characteristic of the anti-$k_T$ algorithm, the $p_T$-seeded method clusters particles based on their actual spatial distribution, and consequently, it generates intuitive results. Our method allows for an easier prediction of the jet position, facilitating straightforward visual identification of the jet's location. By incorporating the $p_T$ seed, the numerous low-$p_T$ particles can be disregarded in the initial round, leading to a substantial enhancement in the compilation speed. Eventually, the $p_T$-seed method can complement the anti-$k_T$ algorithm in the clustering of mini-jets, particularly in the low-$p_T$ region where the application of the anti-$k_T$ algorithm might not be suitable.

\section{Conclusions}
\label{Sec:VII}

\begin{figure}[tbhp]\centering
\includegraphics[width = .3\linewidth]{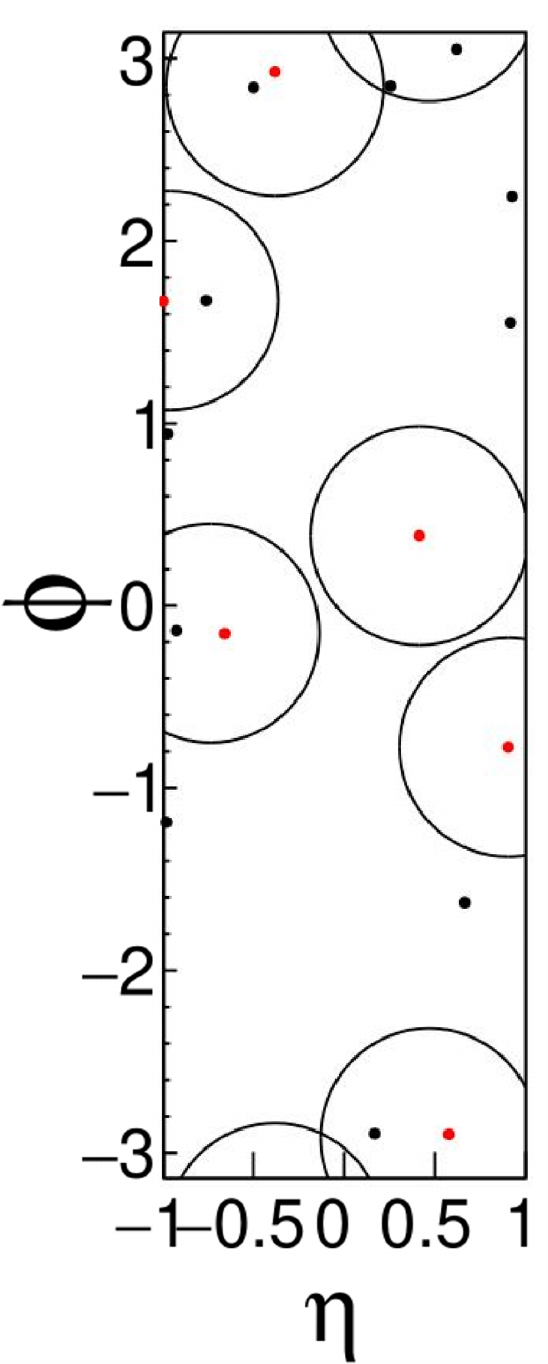}
\caption{An example PYTHIA event with $M$ = 17, $K$ = 6, and $p_{T0} =$ 1 GeV/$c$.}
\label{fig3-1}\end{figure}

We have developed a $p_T$-seeded clustering algorithm, which utilizes a threshold $p_{T0}$ constraint on the seed to exclude low-$p_T$ particles during the initial clustering steps so that jets that are solely composed of low-$p_T$ particles are eliminated from the output. The high-$p_T$ seed particles are merged based on their $\eta$-$\phi$ distance, and subsequently, the clustering process is refined by incorporating low-$p_T$ particles. The design is aimed at minimizing the effects of random cluster noise in the low-$p_T$ range. We have evaluated the performance of our algorithm using a PYTHIA data set. The clustering results indicate the expected hard-scattering-type correlations. Notably, the performance characteristics vary with $p_{T0}$, with a $p_{T0}$ range of 1 to 1.5 GeV/$c$ yielding a clear distinction between the same events and the mixed events.

We are excited about the potential to apply our seeded clustering algorithm to real data to investigate novel phenomena, such as collective flow or other mechanisms beyond hard scattering. In actual hadron-hadron collisions, the number of clusters may be significant, as demonstrated in Fig.~\ref{fig3-1}. This presents an exciting opportunity to explore exotic reaction mechanisms.

\begin{acknowledgments}{
G. W. and H. Z. H. are supported by the U.S. Department of Energy under Grant No. DE-FG02-88ER40424 and by the National Natural Science Foundation of China under Contract No.1835002.  C.Y. W. is supported in part by UT-Battelle, LLC, under Contract No. DE-AC05-00OR22725 with the US Department of Energy (DOE). H. J. is supported by the U.S. Department of Energy under Grant No. DE-FG02-86ER40681. N. Y. is supported by the U.S. Department of Energy under Grant No. DE-
SC0023861.
}
\end{acknowledgments}

\end{document}